\documentclass[letterpaper]{jpconf}

\usepackage{graphicx}

\begin{document}

\title{Deconfinement and chiral transition with the highly improved
staggered quark (HISQ) action}

\author{Alexei Bazavov$^a$ and Peter Petreczky$^b$ 
        [The HotQCD collaboration]\footnote{
The HotQCD Collaboration members are:
A.~Bazavov,
T.~Bhattacharya,
M.~Cheng,
N.H.~Christ,
C.~DeTar,
S.~Gottlieb,
R.~Gupta,
U.M.~Heller,
C.~Jung,
F.~Karsch,
E.~Laermann,
L.~Levkova,
C.~Miao,
R.D.~Mawhinney,
S.~Mukherjee,
P.~Petreczky,
D.~Renfrew,
C.~Schmidt,
R.A.~Soltz,
W.~Soeldner,
R.~Sugar,
D.~Toussaint,
W.~Unger
and
P.~Vranas
}}

\address{$^a$Department of Physics, University of Arizona,
         Tucson, AZ 85721, USA}

\address{$^b$Physics Department, Brookhaven National Laboratory, 
         Upton, NY 11973, USA}

\begin{abstract}

We report on investigations of the chiral and deconfinement aspects
of the finite temperature transition in 2+1 flavor QCD using the
Highly Improved Staggered Quark (HISQ) action on lattices with
temporal extent $N_\tau=6$ and $N_\tau=8$. We have performed the
calculations for physical values of the strange quark mass $m_s$ and 
the light quark masses $m_l=0.2m_s$ and $0.05m_s$. 
Several finite temperature observables, 
including the renormalized Polyakov loop, the renormalized 
chiral condensate and the chiral susceptibility have been calculated.
We also study the fluctuations and correlations of different conserved charges
as well as the trace anomaly at finite temperature.
We compare our findings with previous
calculations that use different improved staggered fermion formulations: asqtad, p4
and stout.
\end{abstract}

\section{Introduction}

Quantum Chromodynamics (QCD) is the theory of strong interactions that
successfully describes a wealth of experimentally observed phenomena.
At high energies the coupling of the theory runs to zero \cite{rev1}, 
making
interactions between the constituents of the theory (quarks and gluons)
weaker, eventually approaching the limit of non-interacting ideal gas.

In the region of low energies (up to a few hundred MeV) the coupling of the theory
is of $O(1)$ (thus justifying the name strong interactions) and
the perturbation theory is not reliable. In this phase of the theory
the degrees of freedom are hadrons, while quarks and gluons are confined
into objects with zero net color charge.

The outlined picture suggests that at some temperature a transition from 
the confined hadronic phase to the deconfined quark-gluon plasma (QGP)
phase takes place \cite{gros81}. 
Creating conditions for achieving QGP has been the
subject of recent experiments at the Relativistic Heavy-Ion Collider (RHIC)
\cite{nagle}
at the Brookhaven National Laboratory and is going to be the goal of the
future heavy-ion program at the Large Hadron Collider (LHC) \cite{salgado}
at CERN.

Since the physics of the hadronic phase and the transition to QGP is 
essentially non-perturbative, lattice QCD techniques have been extensively
applied to its studies \cite{physicstoday}. 
Putting fermions on the lattice leads to the
infamous fermion doubling problem, and several lattice fermion
formulations deal with it in different ways.
Improved staggered fermion formulations are widely used to study 
QCD at non-zero temperatures and densities, see e.g. Ref. \cite{carleton,petr}
for recent reviews, for, at least, two reasons:
they preserve a part of the chiral symmetry
of the continuum QCD which allows one
to study the chiral aspects of the finite temperature transition,
and are relatively inexpensive to simulate numerically 
because due to absence of an additive mass
renormalization the Dirac operator is bounded from below.
However, there are at least two problems with staggered fermion formulation.
The first one is the validity of the rooting procedure, i.e. the way to reduce
the number of tastes (unwanted unphysical degrees of freedom that
result from fermion doubling) from four to one, and the other is 
breaking of the taste symmetry
at finite lattice spacing. The discussion of the validity of rooted 
staggered fermions is presented in Refs. \cite{sharpe06,creutz07}. 
To reduce the taste violations smeared links,
i.e. weighted averages of different paths on the lattice that connect
neighboring points,
 are used in the staggered
Dirac operator and several improved staggered formulations, like
p4, asqtad, stout and HISQ differ in the choice
of the smeared gauge links. The ones in the p4 and asqtad actions
are linear combinations of single links and different staples 
\cite{karsch01,orginos} and therefore are not elements of the SU(3) group.
It is known that projecting the smeared gauge fields onto the
SU(3) group greatly improves the taste symmetry \cite{anna}. The
stout action \cite{fodor05} and the HISQ action implement the
projection of the smeared gauge field onto SU(3) (or simply U(3)) group 
and thus achieve better
taste symmetry at a given lattice spacing. For studying QCD at high
temperature it is important to use discretization schemes which 
improve the quark dispersion relation, thus eliminating the 
tree level ${\cal O}(a^2)$ lattice artifacts in thermodynamic quantities.
The p4 and asqtad actions implement this improvement by introducing
3-link terms in the staggered Dirac operator.
In this paper we
report on exploratory studies of QCD thermodynamics with the
HISQ action which combines the removal of tree level ${\cal O}(a^2)$ lattice artifacts
with the addition of projected smeared links that greatly improve the taste symmetry.
We also compare our results with the previous ones obtained
with the asqtad, p4 and stout actions \cite{milc04,p4eos,hotqcd,eos005,fodor06,fodor09}
at comparable quark masses and lattice spacings 
as well as with the asqtad action at smaller quark masses and larger $N_{\tau}$ 
\cite{hotqcd_progress}.

\section{Action and run parameters}

The Highly Improved Staggered Quark (HISQ) action developed 
by the HPQCD/UKQCD collaboration \cite{Follana:2006rc}
reduces taste symmetry breaking and decreases the splitting
between different pion tastes by a factor of about three 
compared to the
asqtad action. The net result, as recent scaling studies show
\cite{Bazavov:2009wm,MILC_hisq_2009}, 
is that a HISQ ensemble at lattice
spacing $a$ has scaling violations comparable to ones 
in an asqtad ensemble at lattice spacing $2/3a$.

In this exploratory study we used the HISQ action 
in the fermion sector
and the tree-level Symanzik improved gauge action without
the tadpole improvement. The strange quark mass $m_s$ 
was set to its physical value adjusting 
the quantity $\sqrt{2m_K^2-m_\pi^2}=m_{\eta_{s\bar s}}\simeq \sqrt{2B m_s}$ 
to the physical  value 686.57~MeV. Two sets of ensembles have
been generated along the two lines of constant physics
(LCP): $m_l=0.2m_s$ and $m_l=0.05m_s$, where $m_l$ is the $u$ and $d$
quark mass. In the first
set runs were performed on $16^3\times 32$
lattices at zero temperature and $16^3\times 6$ at finite
temperature, and in the second set on 
$32^4$ and $32^3\times 8$ lattices, correspondingly.
To estimate the cutoff effects we have also generated a
set of high-temperature ensembles on $24^3\times6$ lattices along
the $m_l=0.05m_s$ LCP.
The parameters and statistics of the runs are
summarized in Table~\ref{tab_runs}. The molecular dynamics (MD)
trajectories have length of 1 time unit (TU) and
the measurements were performed every 5 TUs at zero
and 10 TUs at finite temperature. Typically, at least
300 TUs were discarded for equilibration at the beginning of the simulations.

\begin{figure}
\includegraphics[width=0.485\textwidth]{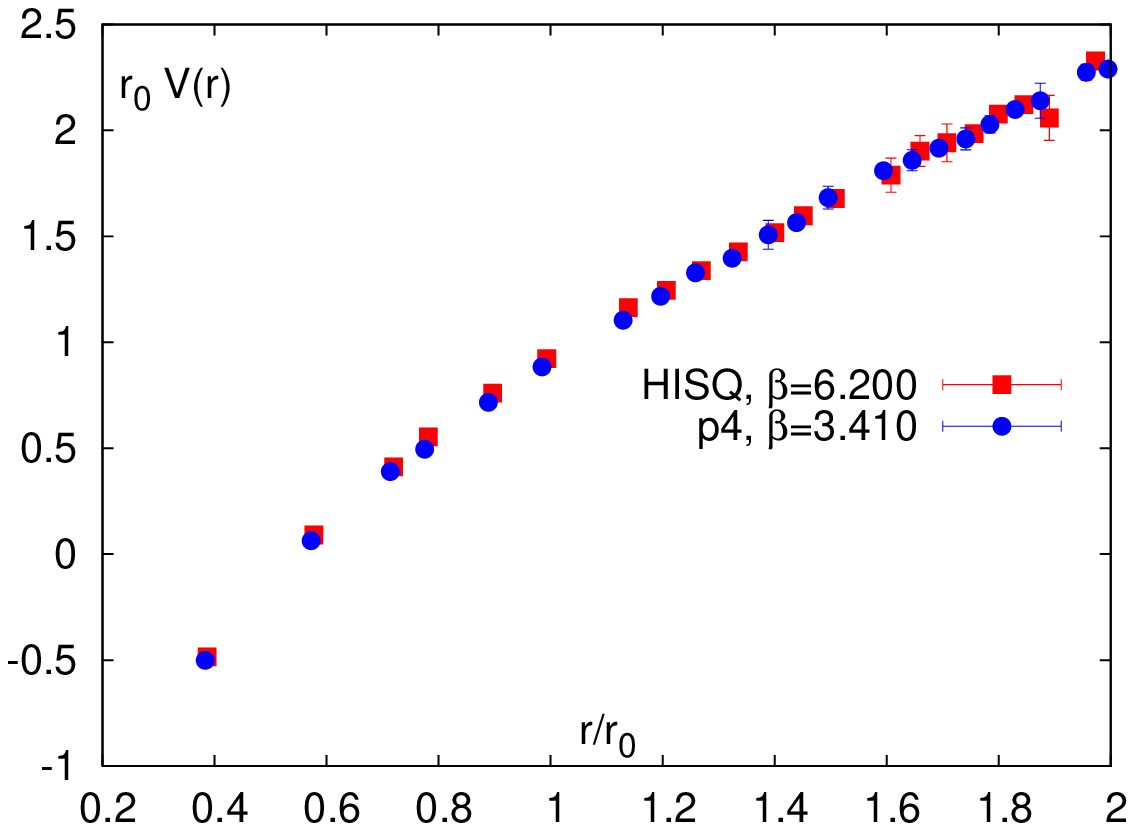}\hfill
\includegraphics[width=0.485\textwidth]{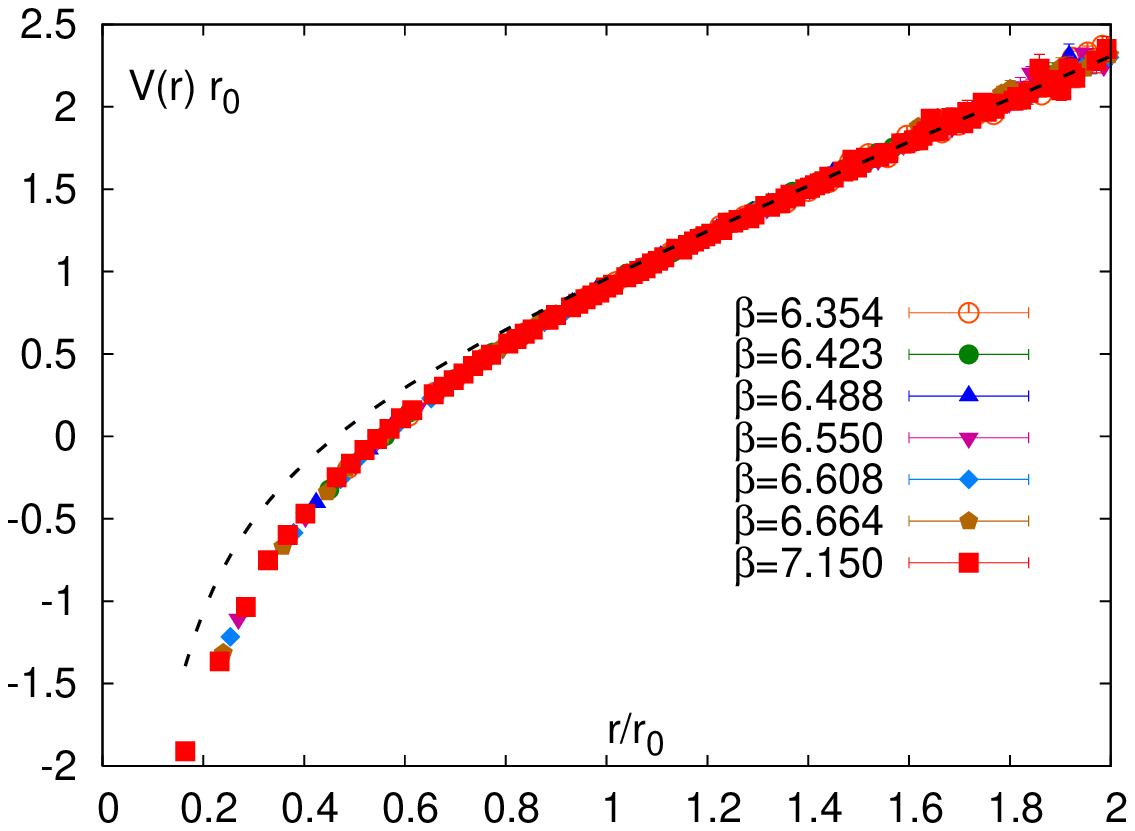}
\caption{The static potential calculated for $m_l=0.2m_s$ (left) and
$m_l=0.05m_s$ (right) in units of $r_0$. In the left figure we compare
the HISQ result with the p4 result obtained at similar value of the lattice
spacing. The dashed line on the right figure is the string potential (see text).}
\label{fig:pot}
\end{figure}

The lattice
spacing has been determined by measuring the static
quark anti-quark potential.
As in previous studies by the MILC collaboration the static potential was calculated fixing the Coulomb gauge
and considering temporal Wilson lines of different extent. Forming the ratio of these correlators
and fitting them to constant plus linear Ansatz we extracted the static potential $V(r)$. We have calculated
the Sommer scale $r_0$ and the related $r_1$ scale defined as
\begin{equation}
\left.r^2 \frac{d V}{dr}\right|_{r=r_0}=1.65,~~~~~
\left.r^2 \frac{d V}{dr}\right|_{r=r_1}=1.00.
\end{equation}
The static potential calculated for $0.2m_s$ and $0.05m_s$ is shown in 
Fig.~\ref{fig:pot}.
The potential has been normalized to the string potential
\begin{equation}
V_{string}=-\frac{\pi}{12r}+\sigma r,
\end{equation}
at $r=1.5r_0$ or equivalently to the value $0.91/r_0$ at $r=r_0$.
The additive constant determined by this normalization is used to
calculate the renormalization constant for the Polyakov loop as will be discussed later.
There is no large cutoff dependence visible in the static potential. Furthermore,
the static potential calculated with the HISQ action is very similar 
to that calculated
with the p4 action. For the ratio of $r_0$ and $r_1$ scales we get
\begin{equation}
r_0/r_1=1.459(3),~m_l=0.2m_s,~~~~~
r_0/r_1=1.481(6),~m_l=0.05m_s.
\end{equation}
To convert from lattice units to physical units we use the value 
$r_0=0.469$ fm determined in \cite{gray}.

\begin{table}
\centering
\caption{The parameters of the numerical simulations:
gauge coupling, strange quark mass and the number of time
units (TU), i.e. the number of MD trajectories for each run. Here
TU, 0 stands for the number of time units for zero-temperature runs,
while TU, $T$ is the number of time units for finite-temperature runs.
The last column gives the number of time units for $24^3\times 6$ runs.}
\begin{tabular}{llllllllrrr}
\br
\multicolumn{5}{c}{
$0.2m_s$ LCP runs
} &
\multicolumn{6}{c}{
$0.05m_s$ LCP runs
} \\ \mr
$\beta$ & $a$, fm & $am_s$ & TU, 0       & TU, $T$ &
$\beta$ & $a$, fm & $am_s$ & TU, 0       & TU, $T$ & TU, $T_6$ \\\mr
6.000   & 0.2297  & 0.115  & 3,000       & 6,000 &
6.195   & 0.1899  & 0.0880 & 2,365       & 6,110 & 11,100 \\
6.038   & 0.2212  & 0.108  & 3,000       & 6,000 &
6.285   & 0.1712  & 0.0790 & 2,300       & 6,190 &  6,750 \\
6.100   & 0.2082  & 0.100  & 3,000       & 6,000 &
6.341   & 0.1612  & 0.0740 &   580       & 7,020 &  6,590 \\
6.167   & 0.1954  & 0.091  & 3,000       & 6,000 &
6.354   & 0.1595  & 0.0728 & 2,295       & 5,990 &  5,990 \\
6.200   & 0.1895  & 0.087  & 3,000       & 6,000 &
6.423   & 0.1475  & 0.0670 & 2,295       & 5,990 &  5,990 \\
6.227   & 0.1848  & 0.084  & 3,000       & 6,000 &
6.488   & 0.1388  & 0.0620 & 2,295       & 5,990 &  8,790 \\
6.256   & 0.1800  & 0.081  & 3,000       & 6,000 &
6.515   & 0.1352  & 0.0604 & 2,045       & 10,100& 10,430 \\
6.285   & 0.1752  & 0.079  & 3,000       & 6,000 &
6.550   & 0.1317  & 0.0582 & 2,295       & 5,990 &  7,270 \\
6.313   & 0.1708  & 0.076  & 3,000       & 6,000 &
6.575   & 0.1278  & 0.0564 & 2,295       & 14,500 & 7,330 \\
6.341   & 0.1665  & 0.074  & 3,000       & 6,000 &
6.608   & 0.1241  & 0.0542 & 2,295       & 5,990 &  6,560 \\
6.369   & 0.1622  & 0.072  & 3,000       & 6,000 &
6.664   & 0.1173  & 0.0514 & 2,295       & 5,990 &  8,230  \\
6.396   & 0.1582  & 0.070  & 3,000       & 6,000 &
6.800   & 0.1047  & 0.0448 & 2,295       & 5,990 &  7,000 \\
6.450   & 0.1505  & 0.068  & 3,000       & 6,000 &
6.950   & 0.0921  & 0.0386 & 2,295       & 5,990 &  7,480 \\
        &         &        &             &       &
7.150   & 0.0770  & 0.0320 & 2,295       & 5,990 &  4,770\\
\br
\end{tabular}
\label{tab_runs}
\end{table}

The masses of several hadrons 
measured on zero-temperature ensembles fall into ranges
summarized in Table~\ref{tab_hadrons}.
They give an idea about the lines of constant physics 
in our calculations. The pion and kaon masses
are most sensitive to the choice of the quark mass. We see that the $0.2m_s$ LCP corresponds
to the lightest pion mass of about $300$~MeV, while the other $0.05m_s$ LCP corresponds
to the lightest pion mass of about $160$~MeV. 
The latter is very close to the physical value of about $140$~MeV.
For this reason we refer to $m_l=0.05m_s$ as the physical quark mass. 
From the table we see that the systematic error in the choice 
of the quark masses is about $4\%$ for the $0.2m_s$ LCP and about
$3\%$ for the $0.05m_s$ LCP.
\begin{table}
\centering
\caption{Ranges of masses (in MeV) of several hadrons
for the two sets of ensembles.}
\begin{tabular}{lrr}
\br
          & $0.2m_s$ LCP & $0.05m_s$ LCP \\\mr
$m_\pi$   & 306-312      & 158-160 \\
$m_K$     & 522-532      & 496-504 \\
$m_\rho$  & 850-883      & 786-800 \\
$m_{K^*}$ & 943-968      & 910-930 \\
$m_\phi$  & 1035-1061    & 1032-1057 \\
$m_N$     & 1130-1183    & 1014-1083 \\
\br
\end{tabular}
\label{tab_hadrons}
\end{table}

\begin{figure}
\includegraphics[width=0.485\textwidth]{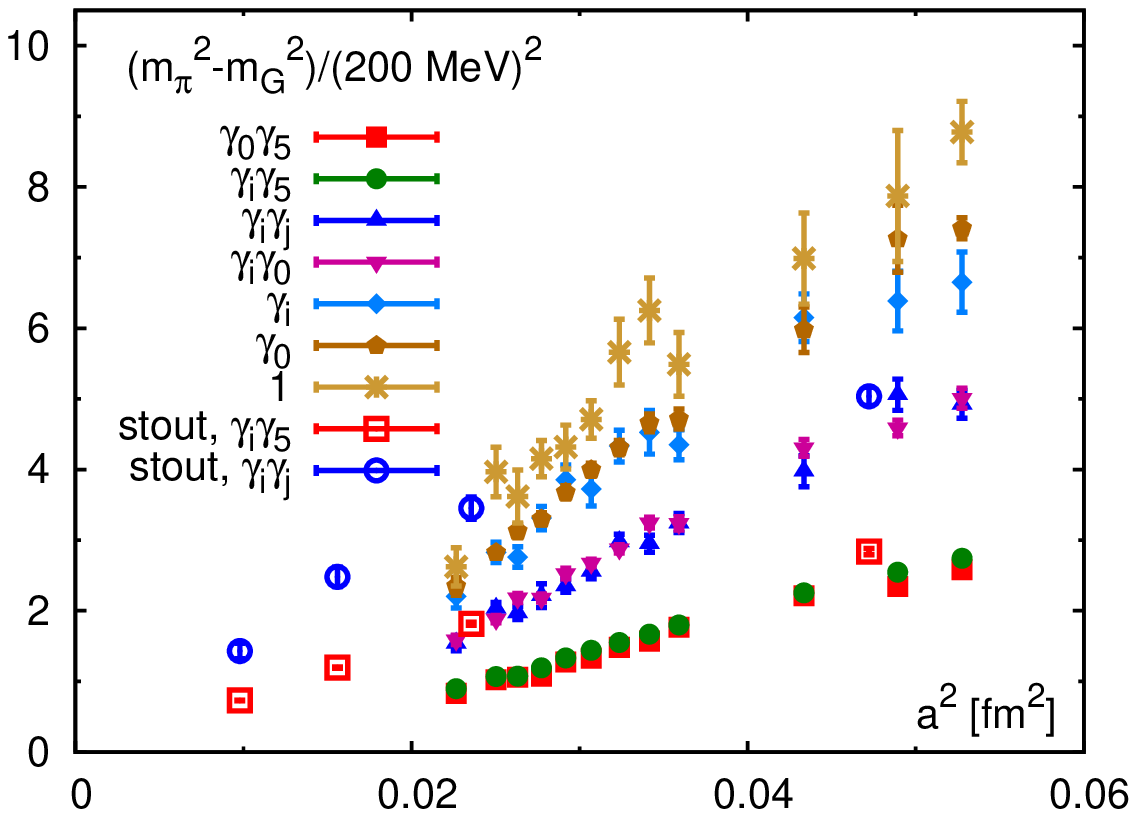}\hfill
\includegraphics[width=0.485\textwidth]{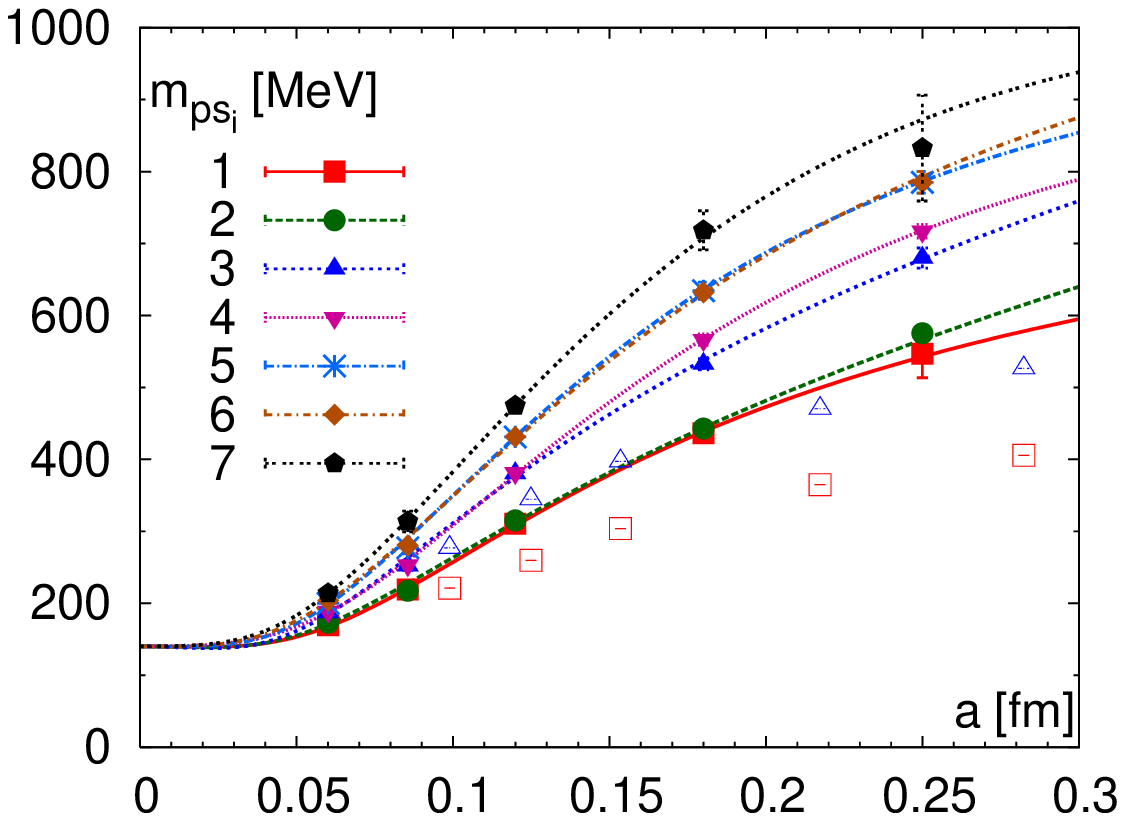}
\caption{The splitting between different pion multiplets for the HISQ action
at $0.2m_s$ compared to the stout results shown as open symbols (left).
In the right figure 
the non-Goldstone pseudo-scalar meson masses are shown as function of lattice spacings assuming that
the lightest (Goldstone) pion mass is fixed to its physical value. The open symbols in the right figure
correspond to pseudo-scalar mesons labeled as ``1'' and ``3'' (or, equivalently, as $\gamma_i \gamma_5$ and 
$\gamma_i \gamma_j$) in the stout calculations. }
\label{fig:psmass}
\end{figure}

As mentioned above the lattice artifacts for the HISQ action are significantly reduced
compared to the asqtad action. The taste violations are strongest in the pseudo-scalar
meson sector. Therefore we studied the splitting of pseudo-scalar meson masses in the
8 different multiplets for $m_l=0.2m_s$. The quadratic splittings of the pseudo-scalar
meson masses are independent of the quark mass to a very good approximation and therefore
it is sufficient to study them for larger value of $m_l$. 
The results are shown in Fig.~\ref{fig:psmass} and compared to
the stout results. In the figure we also show the masses of all the pseudo-scalar mesons
as function of the lattice spacing for asqtad and stout actions assuming that the lightest pseudo-scalar mass 
is fixed to its physical value. For the lattice spacing corresponding to the transition region 
for temporal extent $N_{\tau}=8$ the mass of the heaviest
pion is about $400-600$~MeV for the stout and asqtad action even for the physical values of the light quark mass.
The quadratic splittings in HISQ calculations are 2 to 3 times smaller than in the calculations with
the asqtad action and also somewhat smaller than for the stout action. However, the effect of mass splitting
in the pseudo-scalar sector is non-negligible even for HISQ.

\begin{figure}
\includegraphics[width=0.485\textwidth]{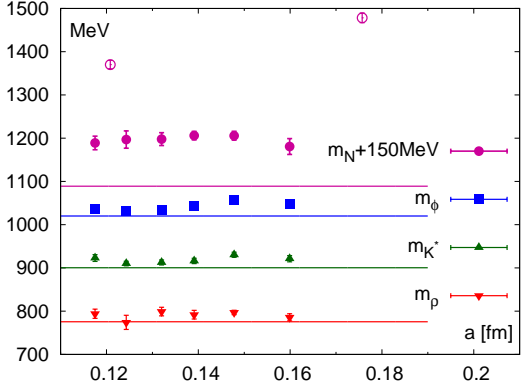}
\begin{minipage}[b]{0.485\textwidth}
\caption{The hadron spectrum for the HISQ action at $m_l=0.05m_s$
compared with experiment. The open symbols show the nucleon mass for
the asqtad action.
For clarity of the plot the nucleon mass is shifted by 150~MeV.
\label{fig:hmass}}
\end{minipage}
\end{figure}

The smaller taste violations also result in better scaling of other hadron masses.
In Fig.~\ref{fig:hmass} the $K^*$, $\rho$ and $\phi$ meson masses as well as the nucleon mass
calculated for $0.05m_s$ are shown.
As one can see there is a good agreement for the vector meson masses 
calculated on the lattice and the experimental values.
The nucleon mass in the HISQ calculations 
is about 10\% larger than the experimental value, but the discrepancy between the
lattice calculations and the experiment is largely reduced compared to the asqtad calculations. 

\section{Deconfinement transition}
The deconfinement transition is usually studied using the renormalized Polyakov loop
\cite{p4eos,hotqcd,fodor06,okacz02}. 
It is related to the free energy of a static quark anti-quark pair at infinite separation~$F_{\infty}(T)$
\begin{equation}
L_{ren}(T)=\exp(-F_{\infty}(T)/(2 T)),
\end{equation}
and obtained from the bare Polyakov loop as
\begin{eqnarray}
&
\displaystyle
L_{ren}(T)=z(\beta)^{N_{\tau}} L_{bare}(\beta)=
z(\beta)^{N_{\tau}} \left<\frac{1}{3}  {\rm Tr } 
\prod_{x_0=0}^{N_{\tau}-1} U_0(x_0,\vec{x})\right >.
\end{eqnarray}
Here the multiplicative renormalization constant $z(\beta)$ is related to the 
additive normalization of the potential $c(\beta)$ as $z(\beta)=\exp(-c(\beta)/2)$
discussed above. To make the comparison with stout data the latter have to be
multiplied by $\exp(-0.91/(r_0 T))$, as in the stout calculation the potential
was normalized to zero at $r=r_0$.

\begin{figure}
\includegraphics[width=0.485\textwidth]{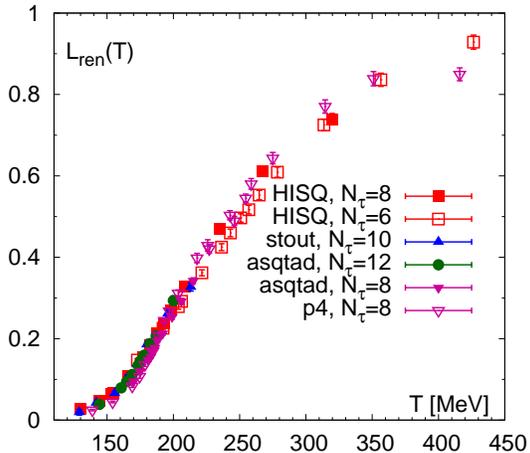}
\begin{minipage}[b]{0.485\textwidth}
\caption{The renormalized Polyakov loop as function of the temperature calculated
for the HISQ, stout, asqtad and p4 actions at the physical value of the
light quark mass.}
\label{fig:poly}
\end{minipage}
\end{figure}

Our results for the quark mass $m_l=0.05m_s$ on $32^3 \times 8$ lattices 
are shown in Fig. \ref{fig:poly} and compared to the stout, asqtad and p4 calculations.
As one can see the HISQ calculations agree reasonably well 
with the stout results if the scale is set by $r_0$ in the stout calculations
\footnote{In what follows we will show the stout results using the temperature scale
set by $r_0$ instead of the kaon decay constant $f_K$. We used the published
values of $r_0$ and $f_K$ in Refs. \cite{fodor06,fodor09} to convert the two scales.}.
On the other hand, the renormalized Polyakov loop calculated with the p4 and asqtad actions
is noticeably smaller at low temperatures. At temperatures $T>200$~MeV we see 
good agreement for different actions.
The decrease of $F_{\infty}(T)$, and thus the increase in the Polyakov loop could be related 
to onset of screening at high temperatures
(e.g. see discussion in Ref. \cite{petr_hp04}). On the other hand, in the 
low-temperature
region the increase of $L_{ren}$ is related to the fact that there are many static-light meson
states that can contribute to the static quark free energy close to the transition temperature, while
far away from the transition temperature it is determined by the binding energy of the lowest static-light mesons.
The large taste symmetry breaking in the static-light meson sector is probably responsible for the discrepancy
between p4 and HISQ results in the low-temperature region. Overall the differences between p4 and HISQ
results are the smallest for the renormalized Polyakov loop as we will see in the following sections.
This is partly due to the fact that the Polyakov loop is purely a gluonic observable. The taste symmetry breaking
associated with light quarks enters only through loops, and the corresponding effect is smaller than for the
quark number susceptibilities, for example.

\section{The chiral transition}

\begin{figure}
\includegraphics[width=0.485\textwidth]{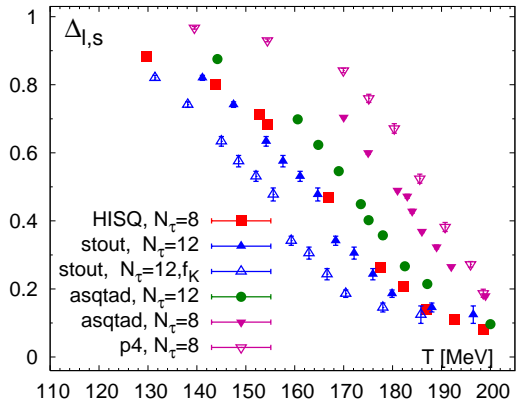}
\begin{minipage}[b]{0.485\textwidth}
\caption{The subtracted chiral condensate for the HISQ action
compared 
with calculations performed for the stout \cite{fodor09}, asqtad \cite{hotqcd_progress} and p4 actions 
\cite{eos005}.
}
\label{fig:pbp}
\end{minipage}
\end{figure}

In the limit of zero light quark masses QCD has a chiral symmetry and
the finite temperature transition is a true phase transition. The order
parameter for this transition is the light chiral condensate 
$\langle \psi \bar \psi \rangle_{l}$.
However, even at finite values of the quark mass the chiral condensate 
will show a rapid change in the
transition region indicating an effective restoration of the chiral symmetry. 
Since the chiral condensate
has an additive ultraviolet renormalization we consider 
the subtracted chiral condensate \cite{p4eos}
\begin{equation}
\displaystyle
\Delta_{l,s}(T)=\frac{\langle \bar\psi \psi \rangle_{l,\tau}-\frac{m_l}{m_s} \langle \bar \psi \psi \rangle_{s,\tau}}
{\langle \bar \psi \psi \rangle_{l,0}-\frac{m_l}{m_s} \langle \bar \psi \psi \rangle_{s,0}}.
\end{equation}
Here the subscripts $l$ and $s$ refer to the light and strange quark condensates 
respectively normalized per single flavor, while the
subscripts $0$ and $\tau$ refer to the zero and finite temperature cases.
In Fig. \ref{fig:pbp} the renormalized chiral condensate calculated 
with the HISQ action is shown
and compared with results obtained with the stout \cite{fodor09},
asqtad \cite{hotqcd_progress} and p4 \cite{eos005} results. 
Our results agree reasonably well with the $N_{\tau}=12$ 
stout results. On the other hand,
the subtracted chiral condensate is considerably smaller than for the asqtad and p4 actions
on $N_{\tau}=8$. This is due
to the larger taste violating effects in the asqtad and p4 case. 
At finer lattice spacings the taste symmetry violations in the asqtad calculations become smaller
and therefore the agreement between the asqtad calculations on $N_{\tau}=12$ lattice
and the HISQ calculations is much better (see Fig. \ref{fig:pbp}).
We also note that if $f_K$ is used to set the scale there is no agreement between HISQ and stout calculations. 

\begin{figure}
\includegraphics[width=0.485\textwidth]{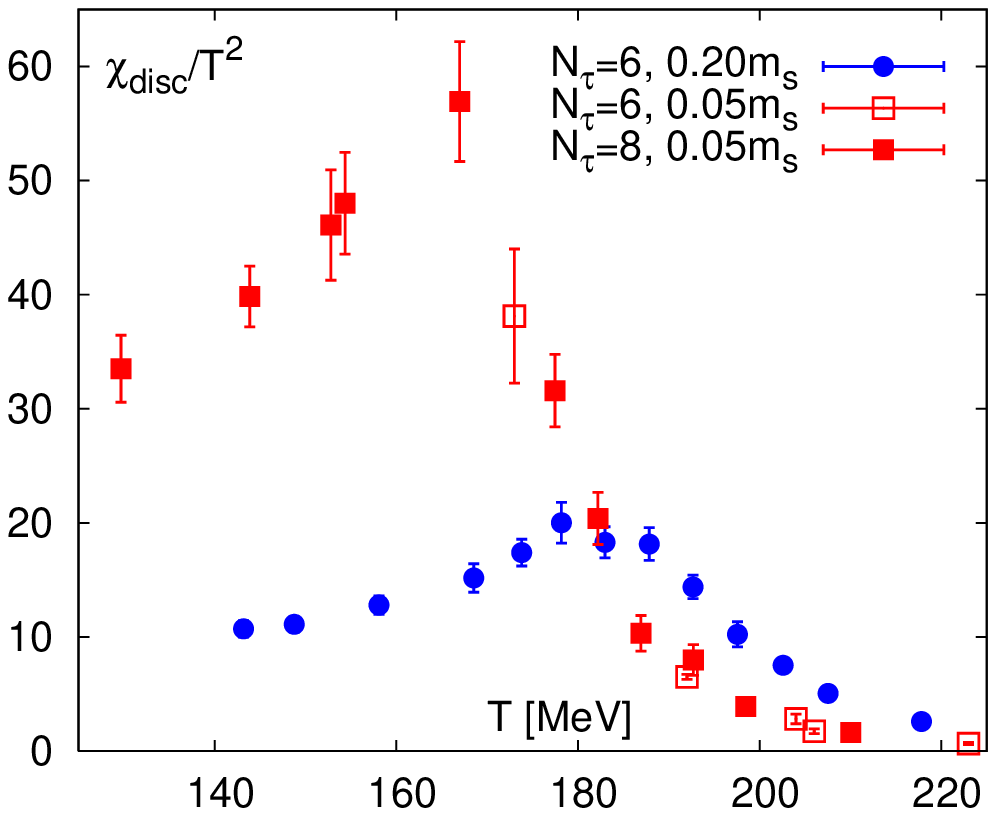}\hfill
\includegraphics[width=0.485\textwidth]{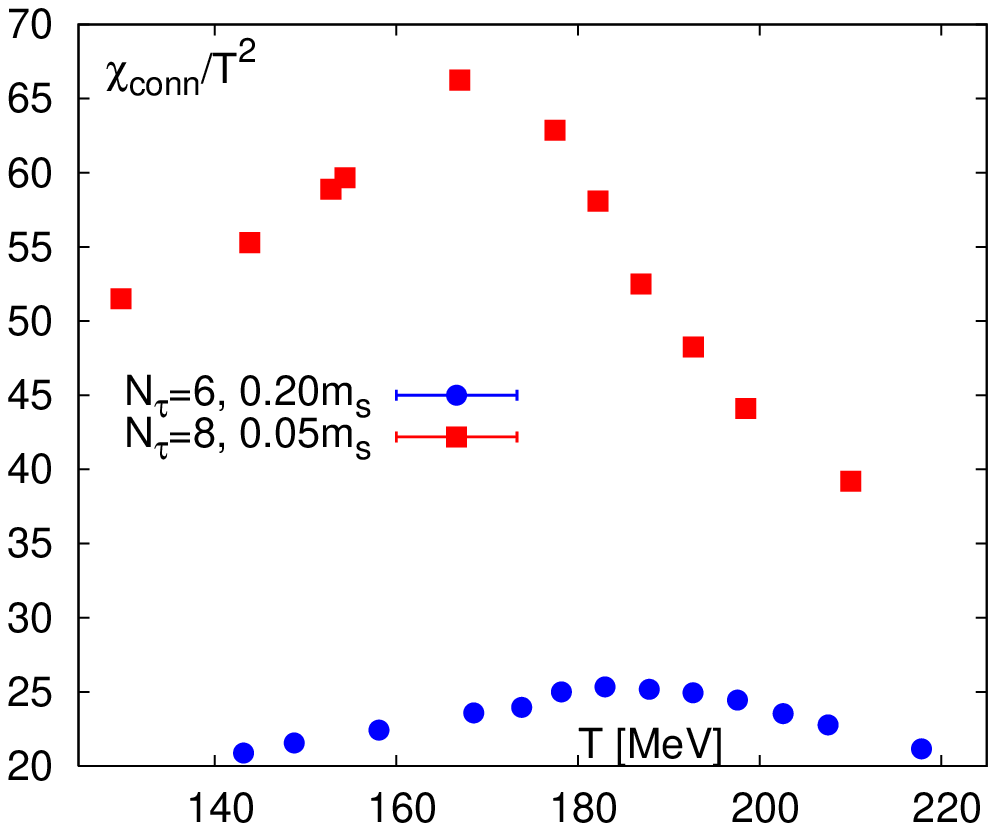}
\caption{
The disconnected (left) and connected (right) chiral susceptibilities calculated for the HISQ action
for the two values of the quark masses. 
}
\label{fig:chi_hisq}
\end{figure}

\begin{figure}
\includegraphics[width=0.485\textwidth]{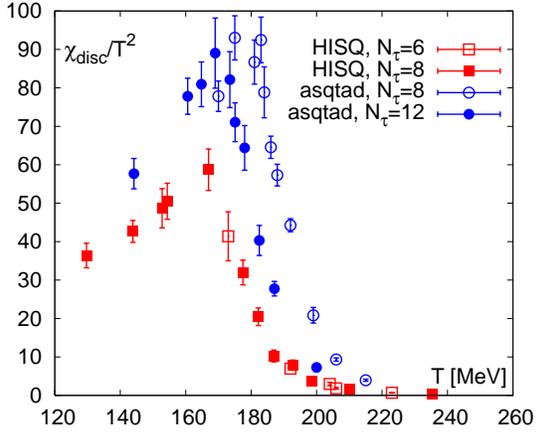}
\begin{minipage}[b]{0.485\textwidth}
\caption{
The comparison of the HISQ and asqtad \cite{hotqcd_progress} results for the disconnected chiral susceptibility
at $m_l=0.05m_s$.
}
\label{fig:chi_comp}
\end{minipage}
\end{figure}

To study the chiral aspects of the QCD transition we also consider the chiral susceptibility defined
as the derivative of the light chiral condensate with respect to the quark mass\footnote{This is the
2-flavor chiral susceptibility as $m_l$ denotes the light quark mass which is common for $u$ and $d$ quarks.}
\begin{equation}
\chi(T)=\frac{\partial \langle \bar\psi \psi \rangle_{l}}{\partial m_l}=\frac{T}{V} \left(
\langle ({\rm Tr} M^{-1}_l)^2 \rangle-\langle{\rm Tr} M^{-1}_l \rangle^2-2 \langle{\rm Tr} M_l^{-2} \rangle \right),
\end{equation}
where $M_l=D+2 m_l$ is the staggered fermion matrix for light quarks. The first term in the above 
formula describes the fluctuation
of the chiral condensate and is also called disconnected chiral susceptibility. 
The second term is called the connected chiral susceptibility and corresponds to the integrated scalar meson
correlation function. For sufficiently small quark masses the chiral susceptibility is dominated by
the disconnected part. The transition temperature is defined as the location
of the peak of the chiral susceptibility.
In Fig. \ref{fig:chi_hisq} the disconnected and connected 
chiral susceptibilities for the HISQ action are displayed.
We see that both of them show a peak-like structure for temperatures 
around $170$ MeV. The disconnected chiral susceptibility has a broad shoulder below the transition
temperature. This is due to the effect of the Goldstone modes \cite{gold}.
In Fig. \ref{fig:chi_comp} we compare the HISQ results for the disconnected chiral susceptibility
with the corresponding results for asqtad calculations performed at $0.05m_s$ \cite{hotqcd_progress}.
There is a good agreement between $N_{\tau}=8$ HISQ and $N_{\tau}=12$ asqtad calculations for the
peak position. 

\begin{figure}
\includegraphics[width=0.485\textwidth]{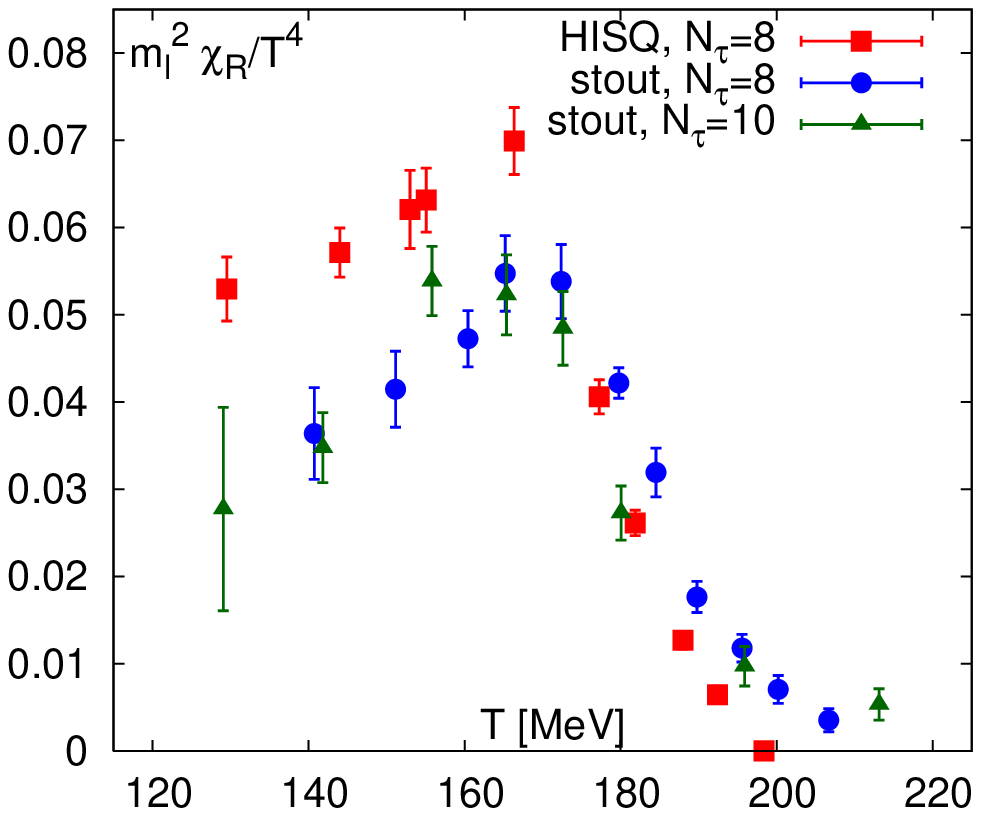}\hfill
\includegraphics[width=0.485\textwidth]{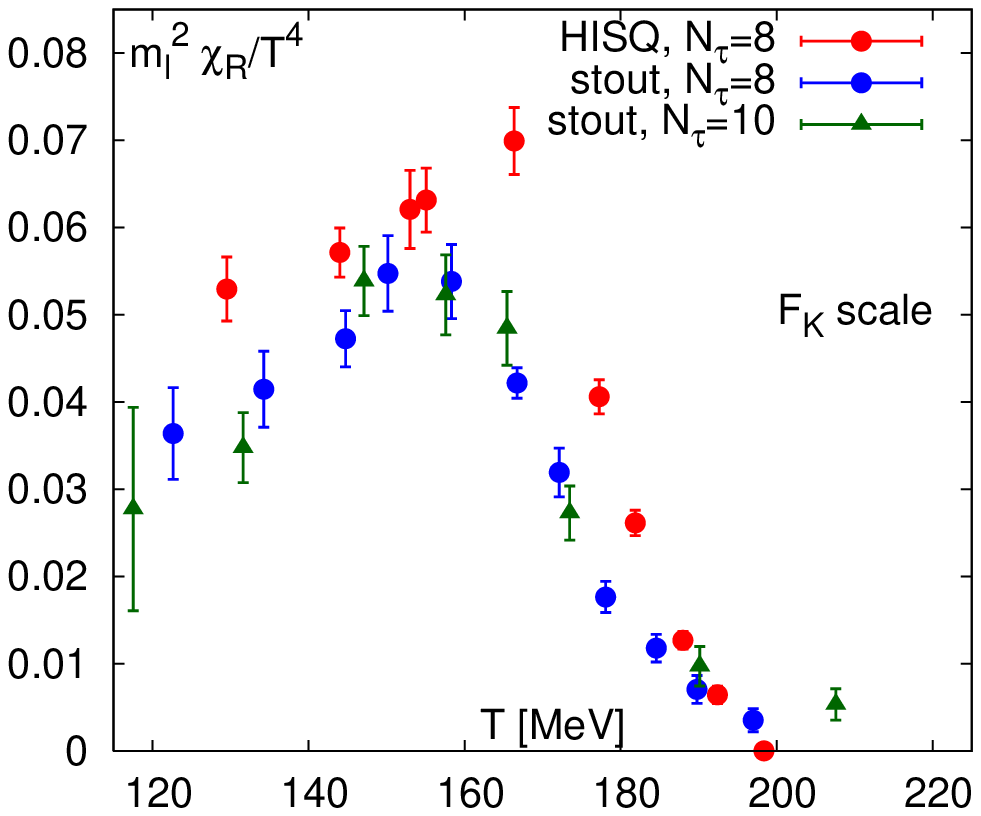}
\caption{
The renormalized chiral susceptibility defined by Eq.~(\ref{chir}) for 
the stout action with lattice
spacing fixed by $r_0$ (left) and $f_K$ (right) \cite{fodor06}. We compare the stout results
with the HISQ calculations at $m_l=0.05m_s$.
}
\label{fig:chi_stout}
\end{figure}

While the disconnected chiral susceptibility is finite in the continuum limit, the connected
part is quadratically divergent for non-vanishing quark masses. To remove this additive ultraviolet divergences
the Budapest-Wuppertal group subracted the zero temperature piece from the chiral susceptibility \cite{fodor06}.
To remove the multiplicative renormalization they also multiplied the chiral condensate by the light 
quark mass squared \cite{fodor06},
i.e. they considered the quantity
\begin{equation}
\chi_r(T)=\frac{m_l^2}{T^4} \chi_R=\frac{m_l^2}{T^4} 
\left(\chi(T)-\chi(T=0)\right).
\label{chir}
\end{equation} 
We calculated the same quantity for the HISQ action 
and show the comparison in Fig. \ref{fig:chi_stout}.
There is a reasonable agreement between HISQ and stout results if the $r_0$ is used to fix the scale. However,
if the $f_K$ is used to set the scale the stout data shift to smaller temperatures.

\section{Fluctuations of conserved charges}

Fluctuations and correlations of conserved charges are good probes of deconfinement because they are sensitive to the underlying degrees of freedom, 
i.e. they can tell whether the relevant
degrees of freedom of the system at a given temperature are hadronic or partonic. Here we consider quadratic
fluctuations and correlations of conserved charges defined as
\begin{eqnarray}
&
\displaystyle
\frac{\chi_i(T)}{T^2}=
\left.\frac{1}{T^3 V}\frac{\partial^2 \ln Z(T,\mu_i)}{\partial (\mu_i/T)^2} \right|_{\mu_i=0},\\
&
\displaystyle
\frac{\chi_{11}^{ij}(T)}{T^2}=
\left.\frac{1}{T^3 V}\frac{\partial^2 \ln Z(T,\mu_i,\mu_j)}{\partial (\mu_i/T) \partial (\mu_j/T)} \right|_{\mu_i=\mu_j=0}.
\end{eqnarray}
Here indices $i$ and $j$ refer to different conserved charges, like quark number, baryon number, strangeness, isospin etc.

\begin{figure}
\includegraphics[width=0.485\textwidth]{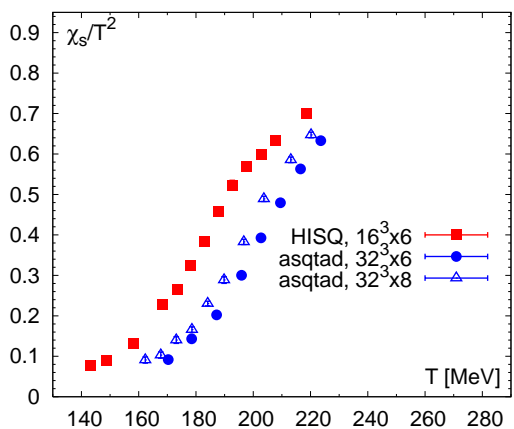}
\begin{minipage}[b]{0.485\textwidth}
\caption{The strangeness fluctuations calculated with the HISQ action and compared to
the asqtad action \cite{hotqcd_progress} for $m_l=0.2m_s$.}
\label{fig:chis02} 
\end{minipage}
\end{figure}

\begin{figure}
\includegraphics[width=0.485\textwidth]{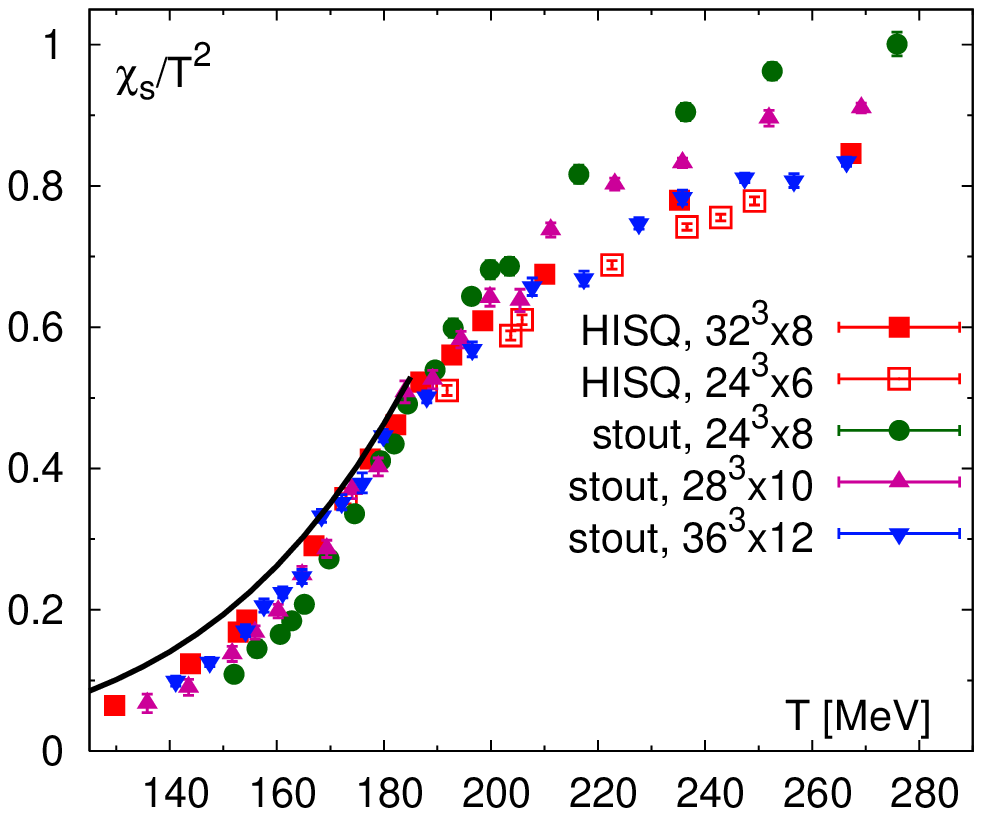}
\hfill
\includegraphics[width=0.485\textwidth]{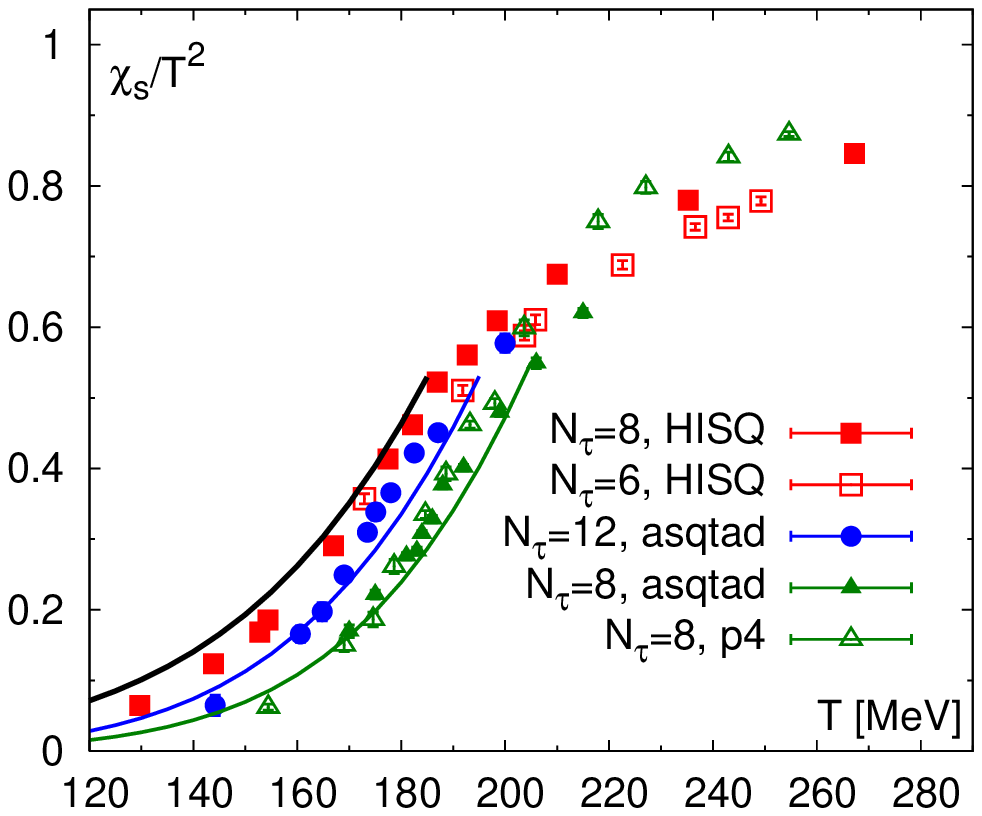}
\caption{The strangeness fluctuations calculated with the HISQ action for $m_l=0.05m_s$ and compared
to the results obtained with the p4, asqtad and stout actions.
The lines show the prediction of the HRG model.
}
\label{fig:chis}
\end{figure}

We start our discussion with the strangeness fluctuation, since this quantity is often discussed in the literature
in connection with deconfinement \cite{hotqcd,fodor06,fodor09}. 
At low temperatures 
strangeness is carried by massive hadrons and therefore its fluctuations are
suppressed. At high temperatures strangeness is carried by quarks and 
the effect of the non-zero strange quark mass is small. Therefore, in the transition region 
the strangeness
fluctuation rises and eventually reaches a value close to that of an ideal quark gas.
For $m_l=0.2m_s$ our numerical results for the strangeness fluctuations 
are shown in Fig. \ref{fig:chis02} and compared with asqtad results.
As one can see the strangeness fluctuations are larger
for the HISQ action at low temperatures, $T<210$ MeV. 
In other words, the transition region in the HISQ calculation shifts toward
smaller temperatures. This behavior is in fact expected. Due to smaller 
taste symmetry violation pseudo-scalar masses as
well as other hadron masses are smaller and therefore 
strangeness fluctuations are larger. 
For the physical quark mass, $m_l=0.05m_s$ we compare our calculations 
with the results obtained using the p4 action \cite{eos005}
and the stout action \cite{fodor06,fodor09}. At low temperatures, $T<200$ MeV 
the HISQ results are significantly larger than the p4
results but are in good agreement with the stout results. 
At high temperatures, $T>200$ MeV the strangeness fluctuations
calculated with the HISQ action are in reasonable agreement 
with the p4 results as well as the $N_{\tau}=12$ stout results. We also
see that in this temperature region the stout results show some cutoff 
($N_{\tau}$) dependence. This is due to the fact
that the tree level ${\cal O}(a^2)$ lattice artifacts are not removed in the stout action.
In the low temperature region we expect that the Hadron Resonance Gas (HRG) model 
gives a reasonably good
description of the thermodynamic quantities, including strangeness fluctuations. 
Therefore in Fig. \ref{fig:chis}
we show the prediction of the HRG model. As one can see all lattice results fall below the HRG model result, although
the difference between the lattice data and HRG is the smallest for the HISQ and stout actions.
The discrepancy between the HRG model and the lattice data becomes larger at smaller temperatures, although the model
is expected to be more reliable there.
As explained in Ref. \cite{pasi} 
this is due to taste symmetry violations leading to the distortion of the hadron spectrum, especially in the pseudo-scalar meson sector. 
Due to non-negligible splitting
in the pseudo-scalar meson masses the contribution of  kaons to strangeness fluctuations appears to be smaller than expected
in the continuum \cite{pasi}. 
As has been discussed in Ref. \cite{pasi} this is mostly due to the discretization errors in the hadron spectrum.
Taking into account the lattice spacing dependence of the hadron masses which enter the HRG calculations it is possible
to get good agreement with lattice data \cite{pasi}.
Therefore in the figure we also show the HRG model calculations with hadron masses
evaluated at lattice spacing corresponding to $N_{\tau}=8$ and $N_{\tau}=12$ lattices using the formulas given in Ref. \cite{pasi}. 
As one can see from the figure there is a very good agreement between this modified HRG calculation and lattice results.
We note again, that for $T<160$ MeV the splitting
in the pseudo-scalar sector is the dominant source of the cutoff effects in the strangeness susceptibility.
Finally we note that, when comparing the lattice results with the prediction of HRG we include all 
resonances up to $2.5$ GeV. Including resonances only up to $2.0$ GeV will not change the strangeness fluctuations significantly, 
but further lowering the mass cutoff of the resonances will decrease the fluctuations for $T>160$ MeV.

\begin{figure}
\includegraphics[width=0.485\textwidth]{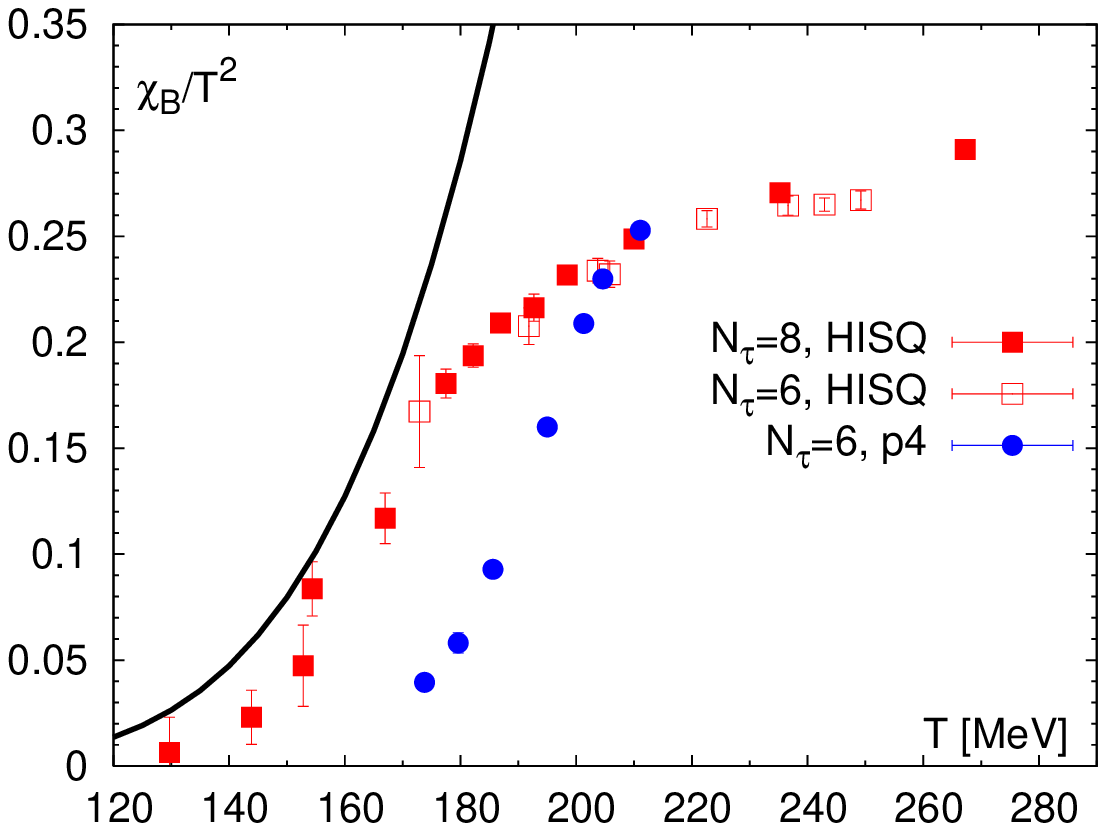}
\includegraphics[width=0.485\textwidth]{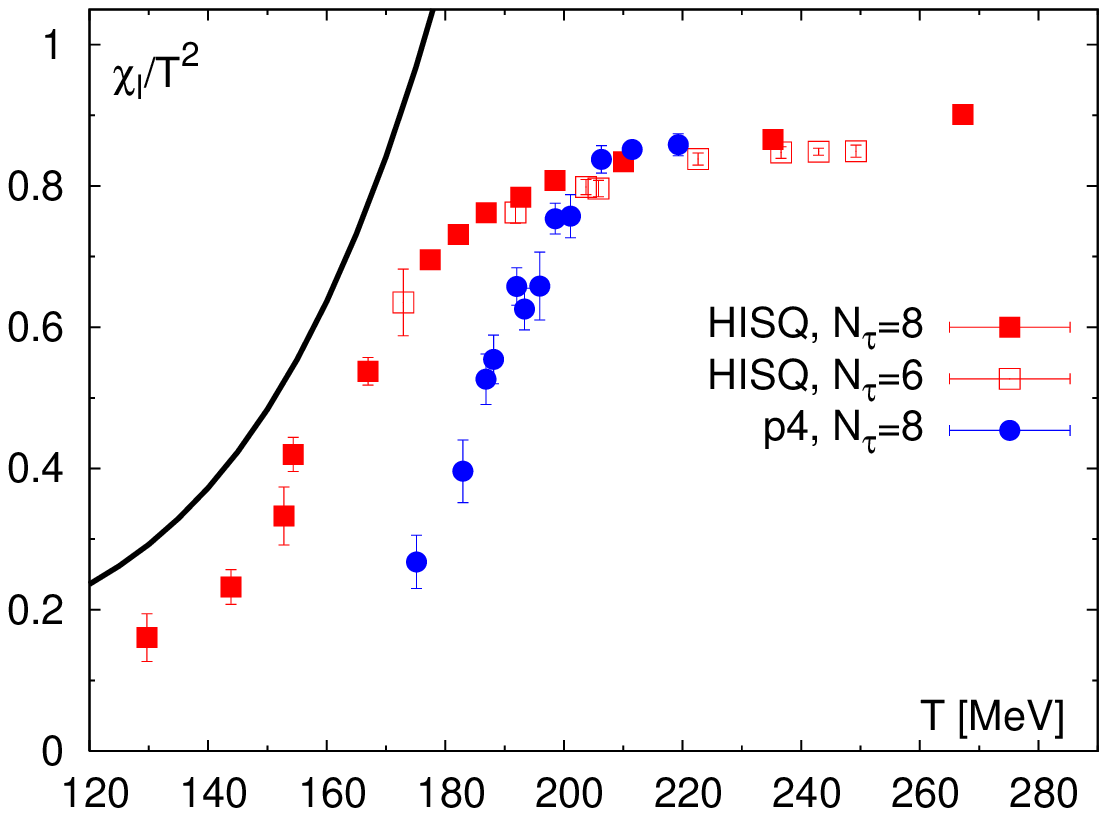}
\caption{Fluctuation of the baryon number (left) and 
quark number (right) calculated with the HISQ action. 
Also shown as the solid line is the prediction of the HRG model. The lattice results are compared with the previous
calculations performed with the p4 action \cite{fluctuations}.}
\label{fig:chiBl}
\end{figure}

We study the quark number and baryon number fluctuations which are sensitive to the light (non-strange) quark or hadron sector. The numerical
results are shown in Fig. \ref{fig:chiBl}. Due to the much smaller quark mass the taste symmetry breaking effects are much
more important in this sector. As the consequence the deviations from the HRG result are larger for these quantities and are significant
even for HISQ calculations.
However,
the deviations between HRG and lattice results obtained with the HISQ action are much reduced compared to the previous calculations
performed with the p4 action on $N_{\tau}=4$ and $6$ lattices at $m_l=0.1m_s$ \cite{fluctuations}. 

\begin{figure}
\includegraphics[width=0.485\textwidth]{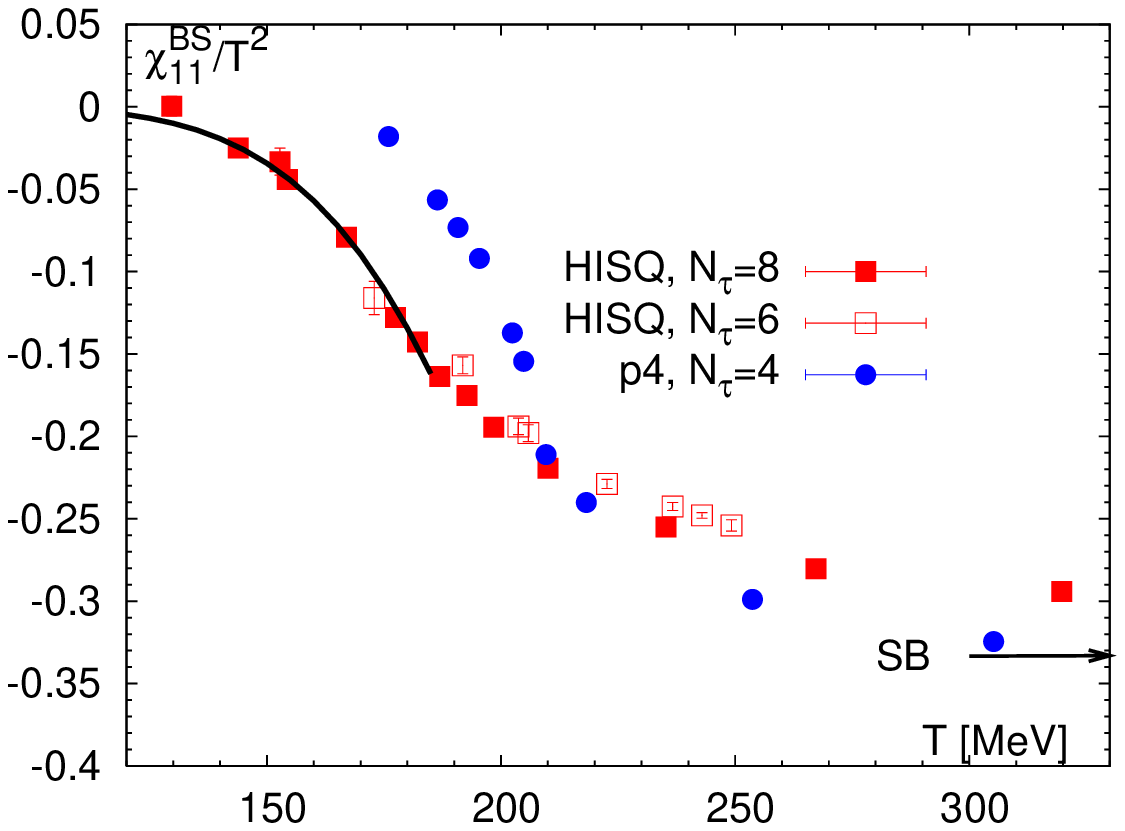}\hfill
\includegraphics[width=0.485\textwidth]{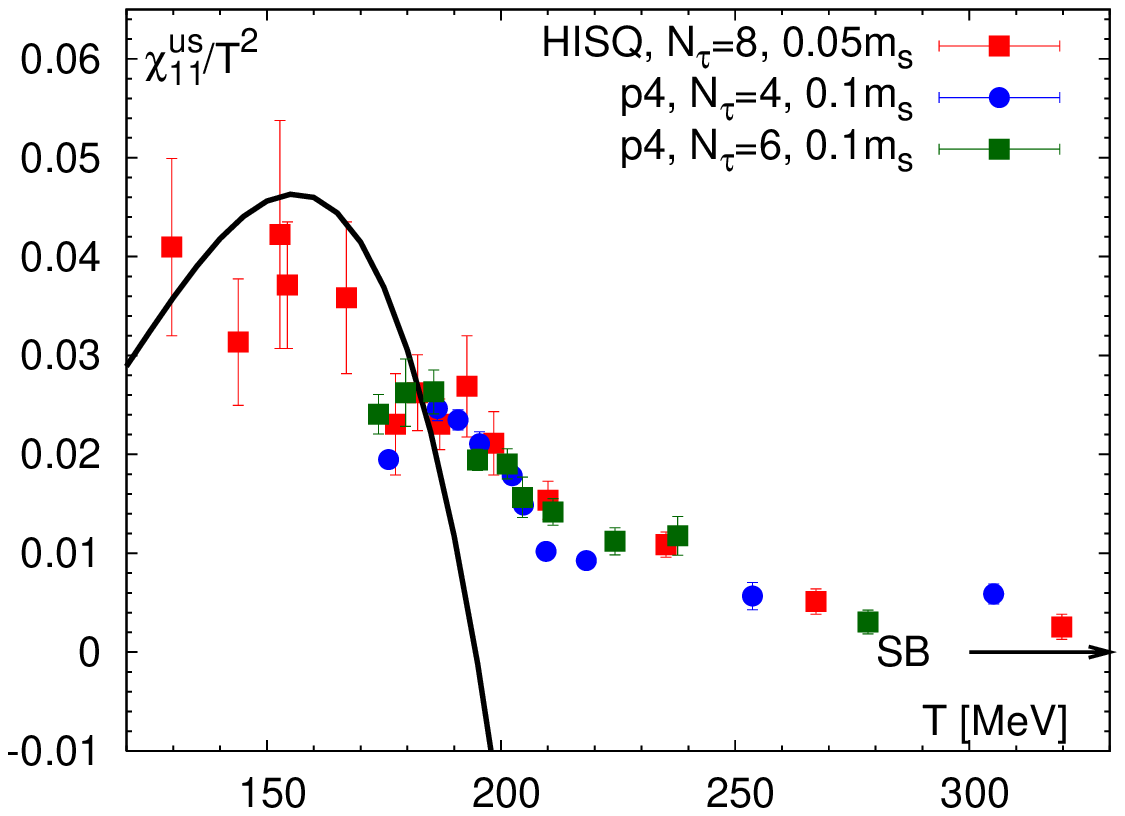}
\caption{Cross-correlation of the baryon number and strangeness (left) and 
and $u$- and $s$-quark number (right) calculated with 
the HISQ action. 
Also shown is the prediction of the HRG model.
The lattice results are compared with the previous
calculations performed with the p4 action \cite{fluctuations}.
}
\label{fig:chi11}
\end{figure}

Next let us consider strangeness-baryon number and $u$- and $s$-quark number correlations. The numerical results
for these quantities are shown in Fig. \ref{fig:chi11}. At low temperatures strangeness-baryon number correlations are due to the presence
of strange baryons. At high temperatures strange quarks are the relevant degrees of freedom and carry $1/3$ unit of the baryon number and
minus one unit of strangeness. As the result strangeness baryon number correlation approaches $-1/3$ at high temperatures. 
The correlations of and $u$- and $s$-quark numbers
at low temperatures are due to the presence of strange hadrons. Here both strange mesons and baryons contribute to this quantity.
At high temperatures quarks
are weakly interacting and the correlation between $u$- and $s$-quark numbers is very small. In fact, in perturbation theory it starts
at order $g^6 \log(g)$ and numerically gives a result which is $10^4$ times smaller than the quark number susceptibility \cite{blaizot}.
At low temperatures correlation of the baryon number and strangeness is very well described by HRG. The lattice results for $u$- and $s$-quark number 
correlations obtained with the HISQ action are not incompatible with HRG, but due to the large statistical errors it is difficult to get more
precise conclusions. The correlations of conserved charges have also been 
calculated with the p4 action, and the corresponding results
are presented in Fig.~\ref{fig:chi11}.
As one can see from the figure the p4 results differ significantly from the HISQ results in the low temperature
region. This is again due to large taste symmetry violations in the p4 calculations.

\begin{figure}
\includegraphics[width=0.485\textwidth]{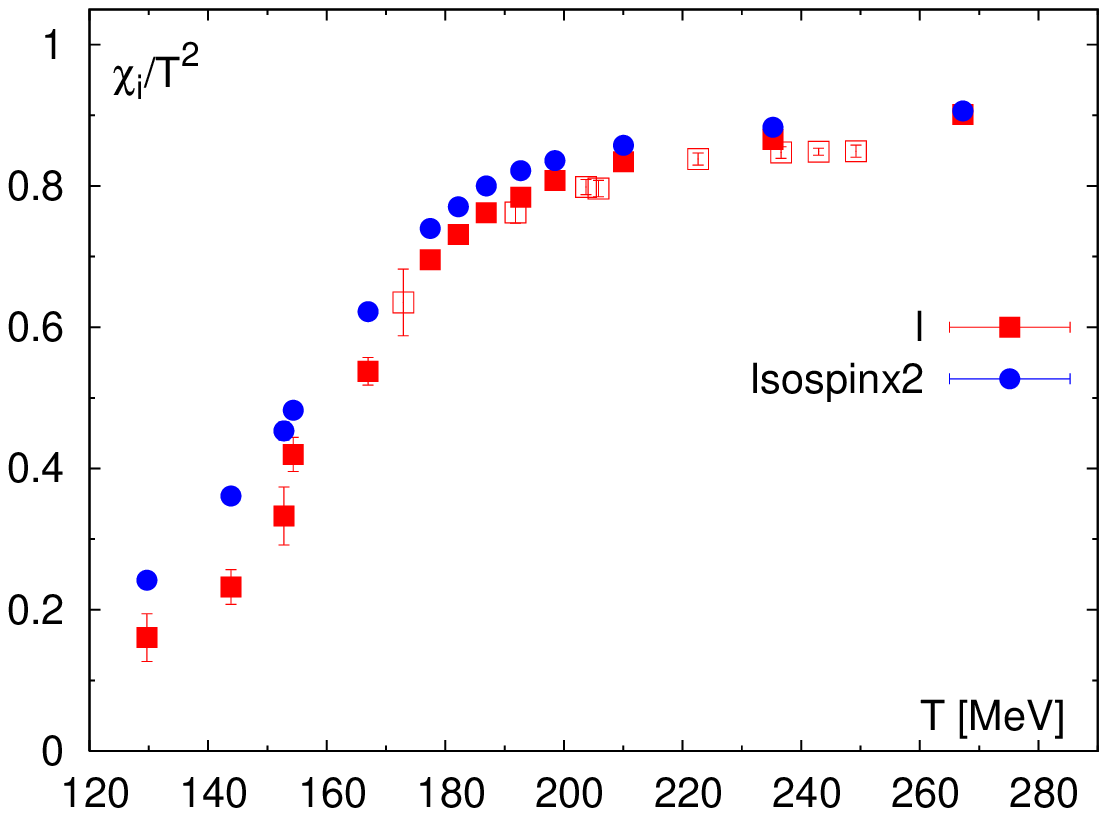}
\includegraphics[width=0.485\textwidth]{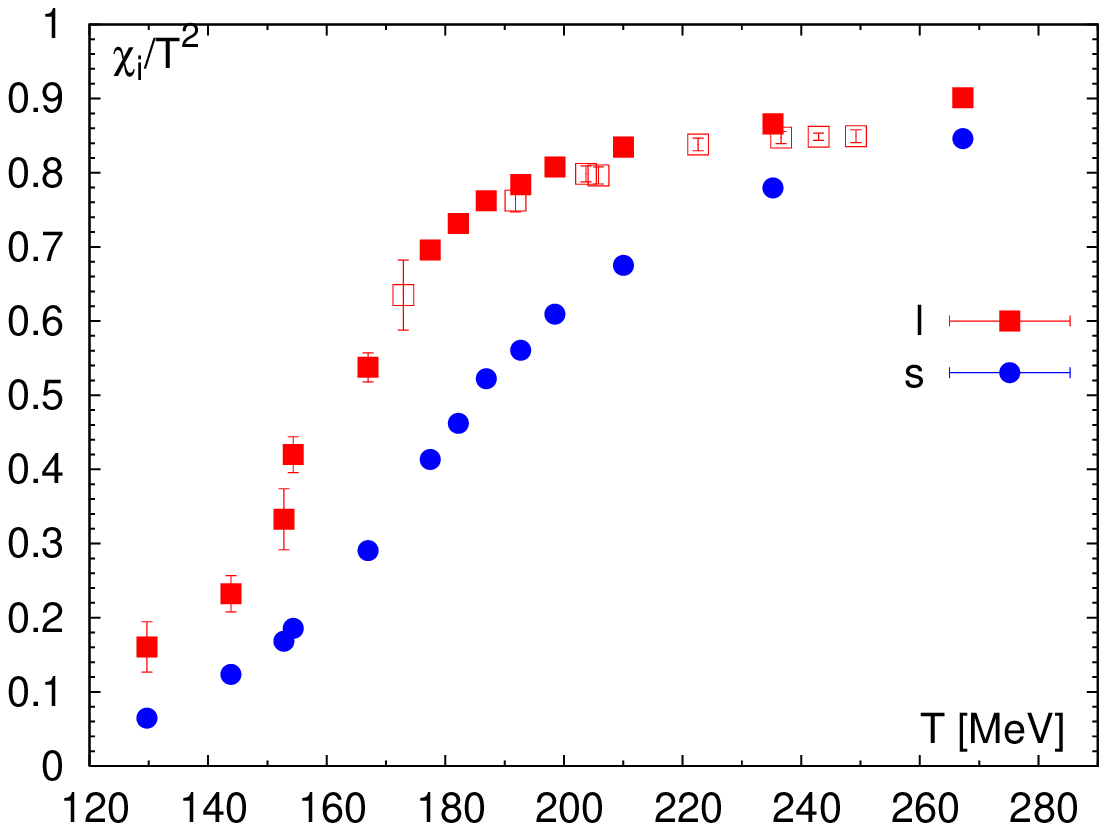}\hfill
\caption{The light quark number susceptibility compared to two times the isospin susceptibility (left) 
and to the strangeness susceptibility (right) for $m_l=0.05m_s$.}
\label{fig:chiIl}
\end{figure}

Finally let us discuss the isospin fluctuations. The numerical results for the isospin
fluctuations are shown in Fig. \ref{fig:chiIl}
and compared with the light quark number fluctuations.
The latter are also compared with the strangeness fluctuations. The isospin fluctuations
are very similar to the light quark number fluctuations at high temperatures. At low temperatures
and in the transition region there are differences between the light quark number and isospin fluctuations.
The strangeness and the light quark number fluctuations are also very different at all temperatures $T<260$~MeV.
Only above this temperature they become similar. One possible reason for the observed  difference could be 
the different sensitivity of these quantities to the singular part of the free energy density. As discussed
in Ref. \cite{hotqcd} the light quark number fluctuations may be sensitive to the singular part of the 
free energy density, while strangeness fluctuations are not sensitive to it. Isospin fluctuations are also
not sensitive to the singular part of the free energy density as they are not coupled to the sigma field \cite{allton}.\footnote{P.P. thanks
F. Karsch and K. Redlich for the discussions on this point.}  It remains to be seen if the differences between the isospin and light quark
number susceptibilities are due to the singular part of the free energy density. Current statistical errors do not allow to make firm conclusions. 

\begin{figure}
\includegraphics[width=0.485\textwidth]{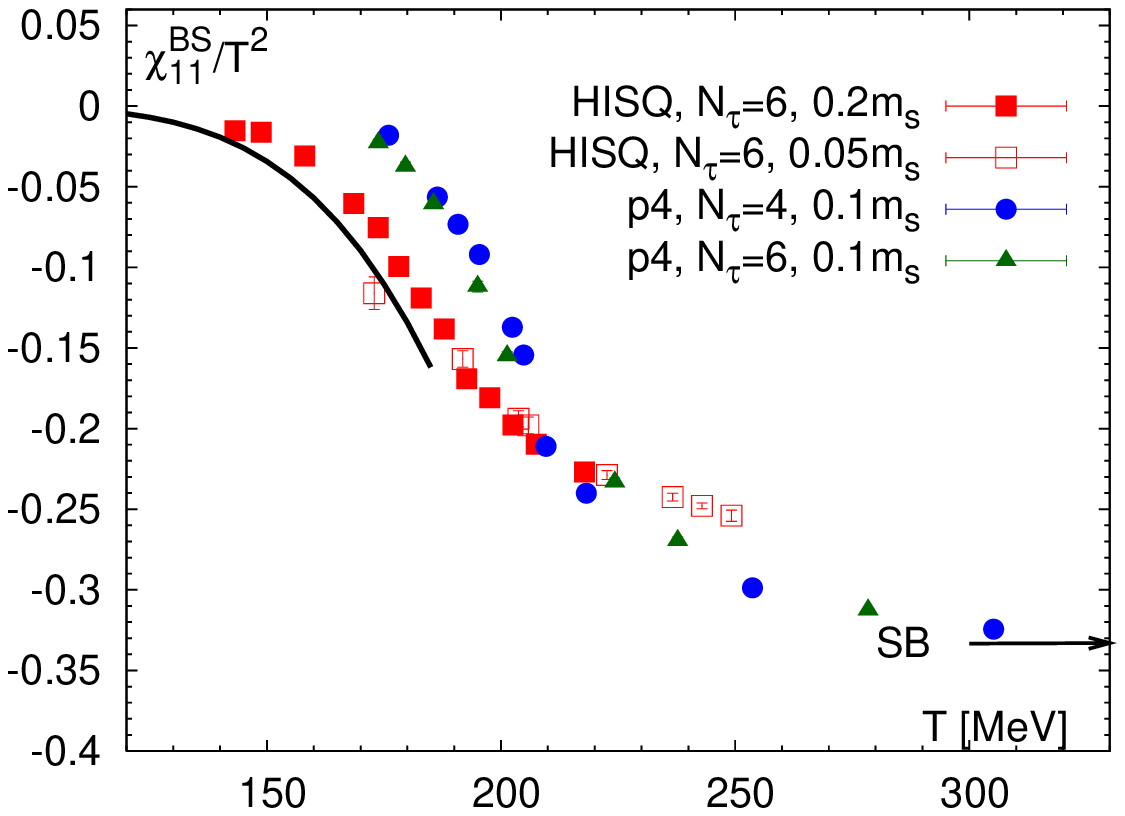}\hfill
\includegraphics[width=0.485\textwidth]{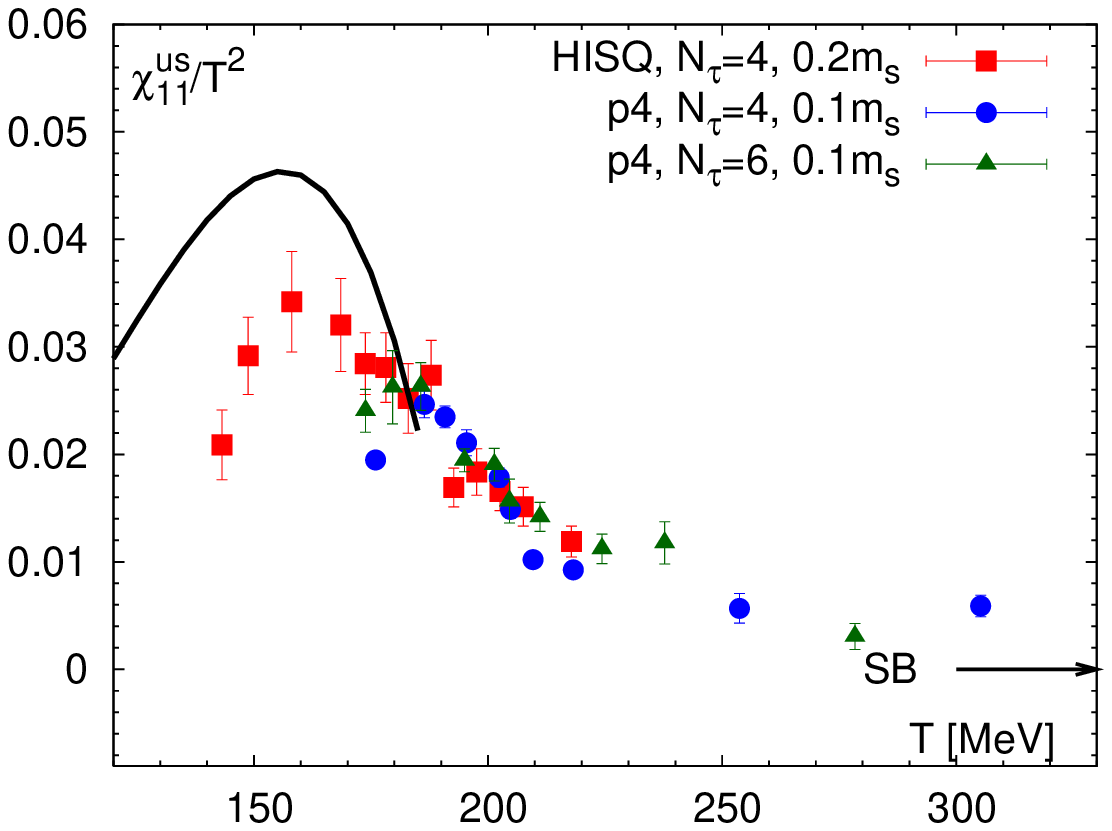}
\caption{Cross-correlation of the baryon number and strangeness (left) and 
and $u$- and $s$-quark number (right) calculated with the HISQ action 
for $m_l=0.2m_s$ and compared with p4 results.
Also shown is the prediction of the HRG model.}
\label{fig:chi11_0.2ms}
\end{figure}

\begin{figure}
\includegraphics[width=0.485\textwidth]{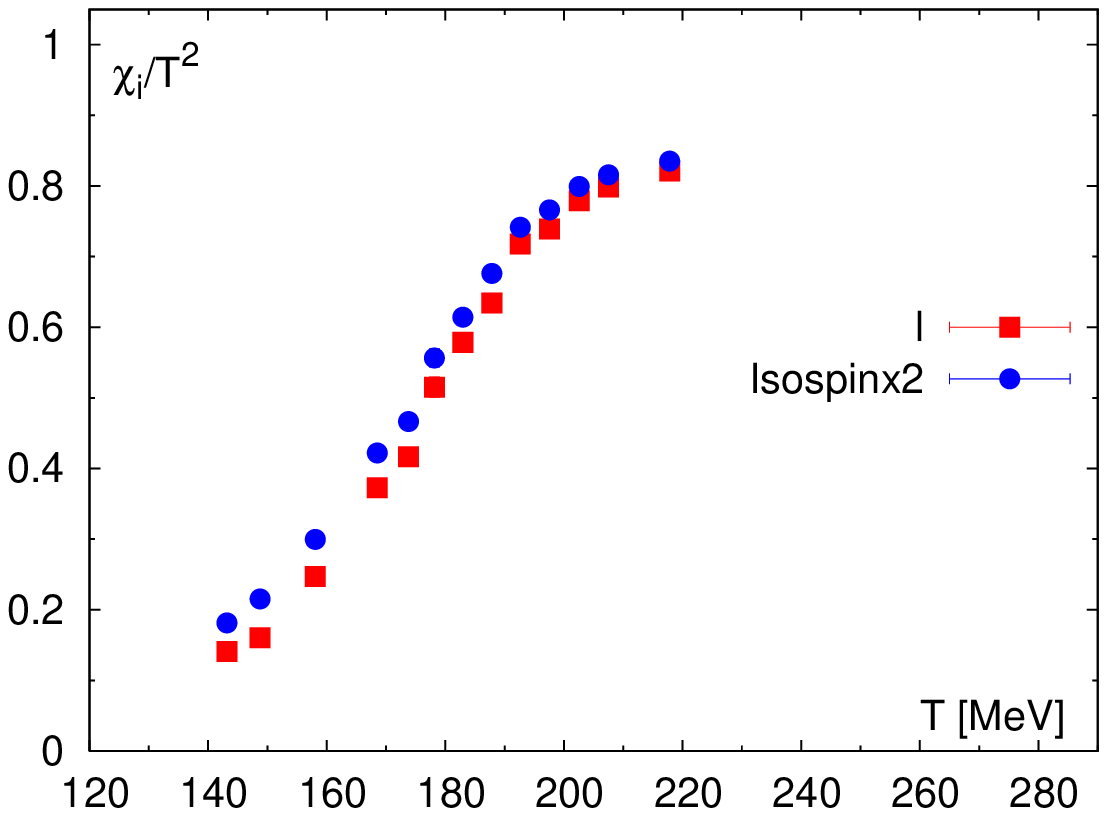}
\includegraphics[width=0.485\textwidth]{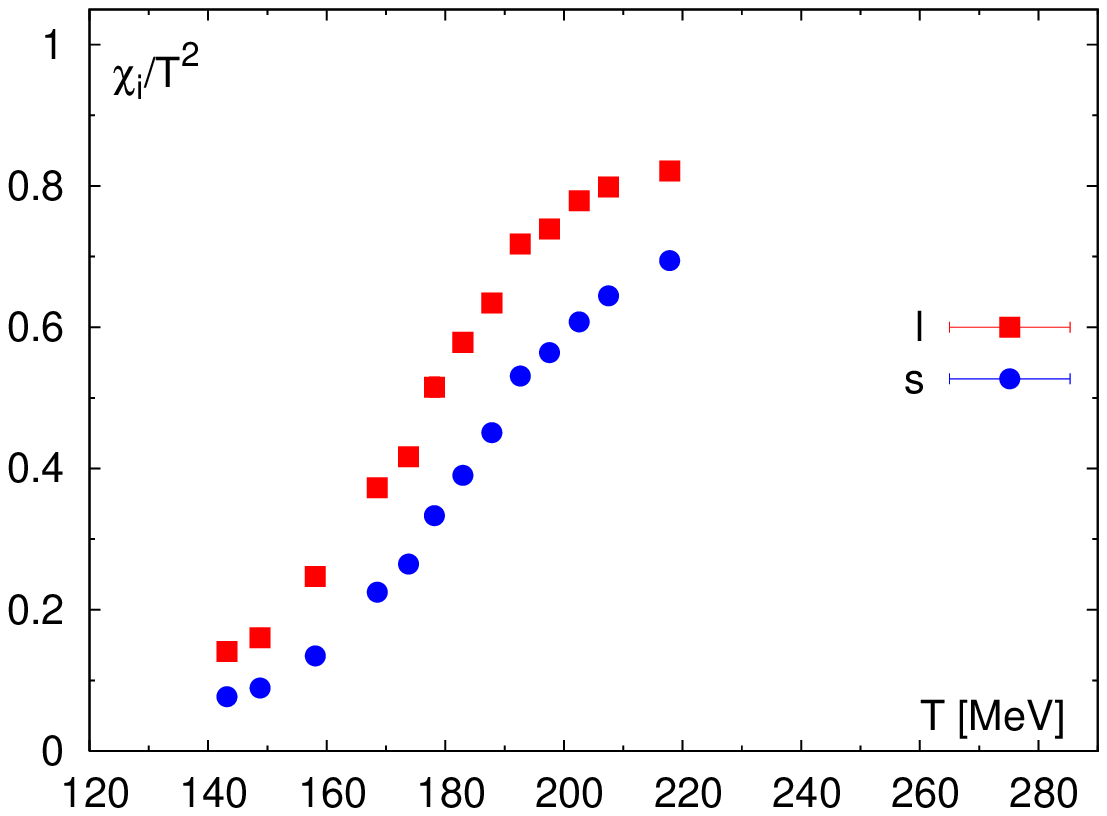}\hfill
\caption{
The light quark number susceptibility compared to two times the isospin susceptibility (left) 
and to the strangeness susceptibility (right) for $m_l=0.2m_s$.}
\label{fig:chiIl_0.2ms}
\end{figure}

Correlations of conserved charges have also been calculated for $m_l=0.2m_s$. The results are shown in Fig. \ref{fig:chi11_0.2ms}.
The results are similar to those obtained for the lighter quark mass. But due to the large quark mass the corresponding 
hadron masses are larger and the agreement with HRG is not that good. In the figure we also compare our numerical results
with the results from previous calculations done with the p4 action \cite{fluctuations}. 
There are significant differences between HISQ and p4 results also for this quark mass. 
Thus cutoff effects are the dominant
source of the difference between  p4 and HISQ results. In Fig.~\ref{fig:chiIl_0.2ms} we show the light quark number
fluctuations in comparison with the isospin and strangeness fluctuations for $m_l=0.2m_s$. As one can see from
the figure the differences between the light quark number and strangeness fluctuations as well as light quark number and isospin
fluctuations are smaller for this quark mass. This could be due to the fact that the differences in the masses of strange and non-strange
hadrons are smaller for $m_l=0.2m_s$, and also due to the reduced sensitivity of the light quark number fluctuations to the singular part. 

\begin{figure}
\includegraphics[width=0.485\textwidth]{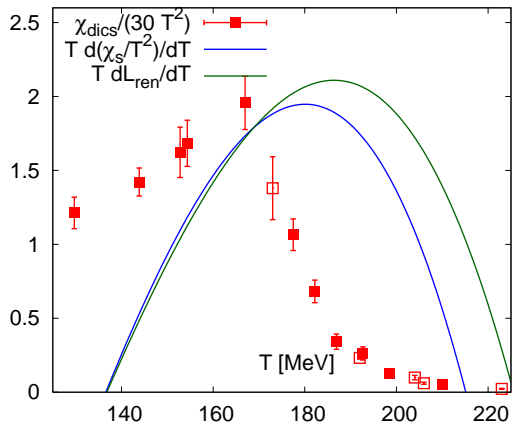}
\begin{minipage}[b]{0.485\textwidth}
\caption{
The comparison of the disconnected chiral susceptibility and the derivatives
of the Polyakov loop and $\chi_s$.
}
\label{fig:joke}
\end{minipage}
\end{figure}

To summarize, in this section we have studied fluctuations and correlations of different conserved charges as function of
temperature. We have found that at low temperatures, $T<170$~MeV, the fluctuations can be well understood in terms of the resonance gas model.
To get a quantitative agreement with HRG it may be necessary to consider modifications of the hadron spectrum due to
finite lattice spacing in the HRG calculations. This has been shown in some detail for the strangeness fluctuations.
At high temperatures, $T>250$~MeV, the fluctuations and correlations can be understood in terms of weakly interacting quark gas.
In the transition region $170\mbox{ MeV}< T < 250\mbox{ MeV}$ there are no well defined quasi-particles which carry the conserved charges, i.e.
it is difficult to define the relevant degrees of freedom.
Thus, the deconfinement transition is a very gradual process if defined as the transition from hadronic to partonic degrees of freedom
and it is impossible to associate a meaningful transition temperature with it. Some fluctuations may be sensitive to the singular part
of the free energy density. In this case the inflection points will be related to the chiral transition temperature.
In the literature (see e.g. Refs. \cite{fodor06,fodor09}) attempts to define the deconfinement transition temperature as inflection points
in the strangeness fluctuations and the renormalized Polyakov loop have been made. From the above discussion it is clear that the deconfinement
transition temperature defined this way has little to do with the deconfinement temperature in the limit of large or infinite quark mass,
which is related to the light modes of the $Z(N)$ center symmetry breaking. Both strangeness fluctuations and the renormalized Polyakov loop
are dominated by the regular part of the free energy density.
Thus, inflection points of these quantities need not to be related to the chiral transition temperature.
In particular, these inflection points could be higher than the chiral transition temperature. This is shown in Fig.~\ref{fig:joke}, where
we compare the disconnected chiral susceptibility with the derivatives of the renormalized Polyakov loop and $\chi_s$. 

\section{Trace anomaly}
\begin{figure}
\includegraphics[width=0.485\textwidth]{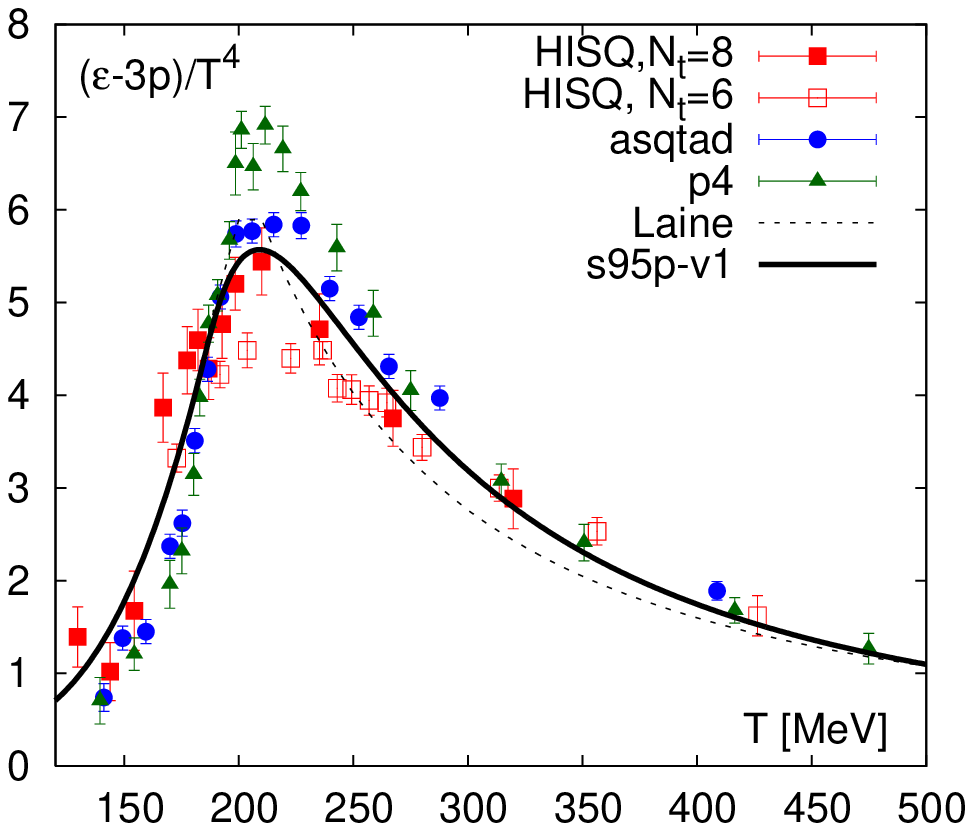}\hfill
\includegraphics[width=0.485\textwidth]{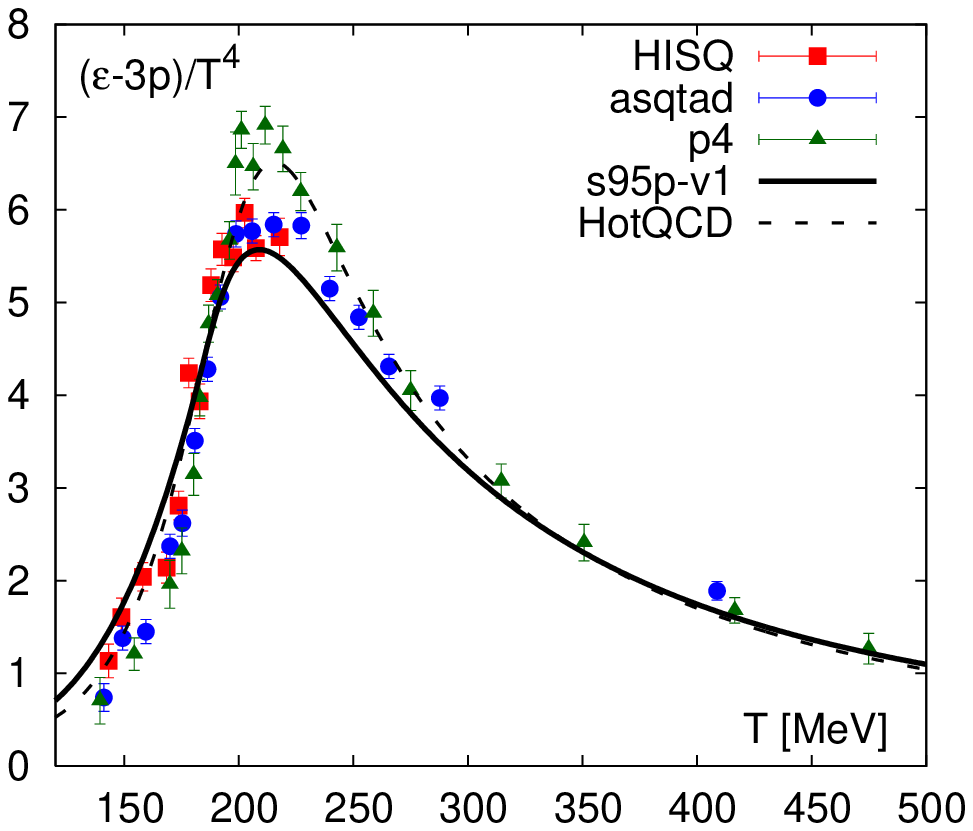}
\caption{The trace anomaly calculated with the HISQ action for $0.05m_s$ (left) and $0.2m_s$ (right)
and compared with p4 and asqtad calculations. The open symbols in the left figure show the 
$N_{\tau}=6$ results. We also compare the lattice results with the parametrizations of the trace
anomaly:  $s95p$-v1 from Ref. \cite{pasi} and the one by Laine and Schroeder \cite{laine}. In the right figure
we also show the HotQCD parametrization of the trace anomaly \cite{hotqcd}.}
\label{fig:e-3p}
\end{figure}

We also calculated the trace anomaly as function of temperature using 
the HISQ action for $m_l=0.2m_s$ and $0.05m_s$.
This quantity is often considered in lattice calculations as  the pressure and other thermodynamic quantities are obtained by integrating
over the trace anomaly to a given temperature.
Previous calculations with the p4 and asqtad action gave a trace
anomaly which was significantly below the HRG result. This deviation from the resonance
gas complicated the use of lattice equation of state in hydrodynamic models since it was not
clear how to implement consistent freezout at the hadronic stage of the evolution (see the discussion
in Ref. \cite{pasi}). In Fig. \ref{fig:e-3p} we show our results for the trace anomaly, $\epsilon-3p$
for the HISQ action and compare them with previous calculations performed with the asqtad and p4 actions.
While at high temperatures ($T>250$~MeV) all lattice data agree, there are noticeable differences
in the low-temperature region. The lattice results obtained with the p4 and asqtad actions are lower
than the HISQ results. This is due to the smaller breaking of the taste symmetry in the calculations performed with the HISQ
action. We also compare the lattice calculations with the parametrization of the trace anomaly obtained from
the matching the HRG at low temperature $T<180$ MeV with the fit to the p4 lattice data at high temperatures
for $T>250$MeV. In the intermediate region the parametrization is constrained by the value of the entropy density
at high temperatures. This
parametrization gives an entropy density which is $5\%$ below the ideal gas value at temperature
of $800$ MeV, and  is called $s95p$-v1 \cite{pasi}.
It was also used in hydrodynamic models \cite{pasi}. 
We also compare the lattice results with the parametrization obtained by Laine and Schroeder \cite{laine}.
Within the statistical errors there is a good agreement between the lattice results obtained 
on $N_{\tau}=8$ lattice with $m_l=0.05m_s$ and the  $s95p$-v1 parametrization. 
In the peak region the trace anomaly calculated on $N_{\tau}=6$ lattice is significantly below the 
$N_{\tau}=8$ results, while at high temperatures the two calculations agree.
Thus, the difference between
$N_{\tau}=6$ and $N_{\tau}=8$ calculations follows the expected pattern of the cutoff dependence of the 
pressure in the free theory, namely at high temperatures the pressure obtained on $N_{\tau}=6$ lattice should be below the
pressure obtained on $N_{\tau}=8$ lattice if the Naik term is used \cite{heller}. We also compare the lattice results with the HotQCD
parametrization of the trace anomaly \cite{hotqcd}. As can be seen in 
Fig.~\ref{fig:e-3p} the HotQCD parametrization agrees with 
$s95p$-v1 parametrization at high temperatures. In the peak region and in the 
low-temperature region the HotQCD and $s95p$-v1 parametrizations are different.
This is due to the fact that the former is based on $N_{\tau}=8$ p4 and asqtad data in the entire temperature range.

\section{Conclusions}

In this paper we discussed recent results 
on QCD thermodynamics with the HISQ action. The use
of this action significantly reduces the effect of the taste symmetry violation in the staggered
fermion formulation. In particular, the quadratic splitting of pseudo-scalar mesons is the smallest
for the HISQ action at a given value of the lattice spacing. The reduced lattice artifacts result 
in more realistic hadron spectrum. 
For instance, the vector meson masses are close to their
continuum values for lattice spacings corresponding to the crossover region 
on $N_{\tau}=8$ lattices along the line of constant physics with
(almost physical) $m_l=0.05m_s$.
We have calculated different thermodynamic quantities with the HISQ action 
on $N_{\tau}=6$ and $N_{\tau}=8$ lattices, including 
the renormalized Polyakov loop, the chiral condensate, chiral susceptibility, 
trace anomaly and the fluctuations
and correlations of conserved charges. 
We have found that at low temperatures the Polyakov loop agrees well with
the results obtained with the stout action, if the
lattice spacing is set by the Sommer scale $r_0$,
while at high temperatures it agrees with the p4 and asqtad calculations. 
The renormalized chiral condensate and chiral susceptibility obtained
with the HISQ action agree quite well with the stout calculations 
but disagree with the results obtained with the p4 and asqtad 
actions on $N_{\tau}=8$ lattices. 
However, the asqtad results on $N_{\tau}=12$ lattices agree well with the ones obtained
with the HISQ action on $N_\tau=8$.

We studied in detail different fluctuations and correlations of conserved charges. When comparing our results 
with previous ones obtained with the p4 and asqtad actions 
we have found differences in the low-temperature region.
Those differences are due to the reduced taste symmetry violation 
in the HISQ action which gives a more realistic 
hadron spectrum. 
We have compared the lattice results in the low-temperature region with the prediction of the hadron
resonance gas. 
We have found good agreement between the lattice results and the HRG once the cutoff effects have been
taken into account. We have also found that the agreement between the lattice 
results and the physical HRG is best for the HISQ action.
The deconfinement transition can be understood as the transition 
from hadronic to quark degrees of freedom as seen
by the different fluctuations/correlations. We have found that this transition is not abrupt but rather gradual. 
This implies that it is not possible to define a meaningful deconfinement transition temperature.

Finally we considered the trace anomaly.  The calculations performed with the HISQ action 
give results which agree
quite well with the previous results obtained with the p4 and asqtad actions 
on $N_{\tau}=6$ and $N_{\tau}=8$ lattices
at high temperatures.
Differences are seen in the transition and in the low-temperature regions. 
The HISQ results on the trace anomaly
agree well with the HRG in the low temperature region, 
and overall they agree with the $s95p$-v1 parametrization that
has been recently proposed for use in hydrodynamic models \cite{pasi}.

\section*{Acknowledgements}
 This work has been supported in part by contracts DE-AC02-98CH10886
 and DE-FC02-06ER-41439 with the U.S. Department of Energy
 and contract 0555397 with the National Science Foundation. The numerical calculations have been performed
 using the USQCD resources at Fermilab as well as the BlueGene/L
 at the New York Center for Computational Sciences (NYCCS).
 We thank Z.~Fodor and S.~Katz for sending us the stout data.

\end{document}